\documentclass{jfm}

\usepackage{natbib}
\usepackage{amsmath,amssymb,mathrsfs,mathtools}
\usepackage{xcolor}
\usepackage{graphicx}
\usepackage{subfigure}
\usepackage{soul}
\usepackage{bbding}
\usepackage[normalem]{ulem}
\usepackage[version=4]{mhchem}
\usepackage[colorlinks=true,urlcolor=blue,linkcolor=blue,citecolor=blue,filecolor=blue]{hyperref}

\usepackage{siunitx}

\usepackage[inline]{trackchanges}
\addeditor{LK}
\addeditor{JD}
\addeditor{TSL}

\usepackage{multirow}

\ifCUPmtlplainloaded \else
  \checkfont{eurm10}
  \iffontfound
    \IfFileExists{upmath.sty}
      {\typeout{^^JFound AMS Euler Roman fonts on the system,
                   using the 'upmath' package.^^J}%
       \usepackage{upmath}}
      {\typeout{^^JFound AMS Euler Roman fonts on the system, but you
                   dont seem to have the}%
       \typeout{'upmath' package installed. JFM.cls can take advantage
                 of these fonts,^^Jif you use 'upmath' package.^^J}%
      }
  \else
  \fi
\fi

\ifCUPmtlplainloaded \else
  \checkfont{msam10}
  \iffontfound
    \IfFileExists{amssymb.sty}
      {\typeout{^^JFound AMS Symbol fonts on the system, using the
                'amssymb' package.^^J}%
       \usepackage{amssymb}%

      }{}
  \fi
\fi

\ifCUPmtlplainloaded \else
  \IfFileExists{amsbsy.sty}
    {\typeout{^^JFound the 'amsbsy' package on the system, using it.^^J}%
     \usepackage{amsbsy}}
    {}
\fi

\newsavebox{\astrutbox}
\sbox{\astrutbox}{\rule[-5pt]{0pt}{20pt}}

\newcommand\nc{\newcommand}
\nc{\vect}[1]{\mbox{\boldmath $#1$}}
\nc{\pt}{\partial_t}
\nc{\px}{\partial_x}

\DeclarePairedDelimiter{\floor}{\lfloor}{\rfloor}

\title[Instabilities of a thin liquid film in a funnel]{Thin liquid films in a funnel}

\author[T.-S. Lin, J.A. Dijksman and L. Kondic]%
{T.-S. Lin$^1$, \ns
and J.A. Dijksman$^2$, \ns
and L. Kondic$^3$
}

\affiliation{
$^1$Department of Applied Mathematics, National Yang Ming Chiao Tung University,
Hsinchu 30010, Taiwan\\[\affilskip]
$^2$Physical Chemistry and Soft Matter, Wageningen University \& Research, Stippeneng 4, 6708WE Wageningen, The Netherlands\\[\affilskip]
$^3$Department of Mathematical Sciences and Center for Applied Mathematics and Statistics, New Jersey Institute of Technology, Newark, New Jersey 07102, USA
}

\pubyear{2014}
\volume{000}
\pagerange{000--000}
\date{?; revised ?; accepted ?. - To be entered by editorial office}

\begin{document}

\maketitle
\begin{abstract}
We explore flow of a completely wetting fluid in a funnel, with particular focus on  contact line instabilities at the fluid front.
While the flow in a funnel may be related to a number of other flow configurations as limiting cases, understanding its 
stability is complicated due to the presence of additional azimuthal curvature, as well as due to 
convergent flow effects imposed by the geometry.  Convergent nature of the flow leads to thickening
of the film, therefore influencing its stability properties. In this work, we analyze these 
stability properties by combining physical experiments, asymptotic modeling, self-similar type of analysis 
and numerical simulations.  We show that appropriate long-wave based model supported by
the input from experiments, simulations and linear stability analysis origination from the 
flow down an incline plane provides a basic insight allowing to understand the development of 
contact line instability and emerging lengthscales.  
\end{abstract}

\section{Introduction\label{sec:intro}}

Thin liquid films with fronts involving contact lines and their instabilities are relevant to applications in a 
number of different fields, ranging from nanoscale to macroscale films where instabilities are driven by a 
combination of various body forces, surface tension, and wettability, see~\citet{oron_rmp97,cm_rmp09} for reviews. 
Significant progress has been reached by using long wave approximation, which simplifies considerably the analysis of 
thin film flows and their stability.
In the context of thin films on a macroscale, the setup involving completely wetting film of 
constant thickness flowing down an incline has been understood reasonably well.   
For such configuration, linear
stability analysis (LSA) carried out in a moving reference frame leads to 
the dispersion relation which shows stability for large wavenumbers, and predicts the most unstable
wavelength (specifying the distance between emerging fingers), which results from the balance between 
destabilizing gravity and stabilizing surface tension forces, see e.g.~\citet{Troian, BB97}.  
However, as soon as some of the simplifying assumptions are removed, understanding the instability becomes 
much more complicated. 

\begin{figure}
\centering
\includegraphics[scale=1.1]{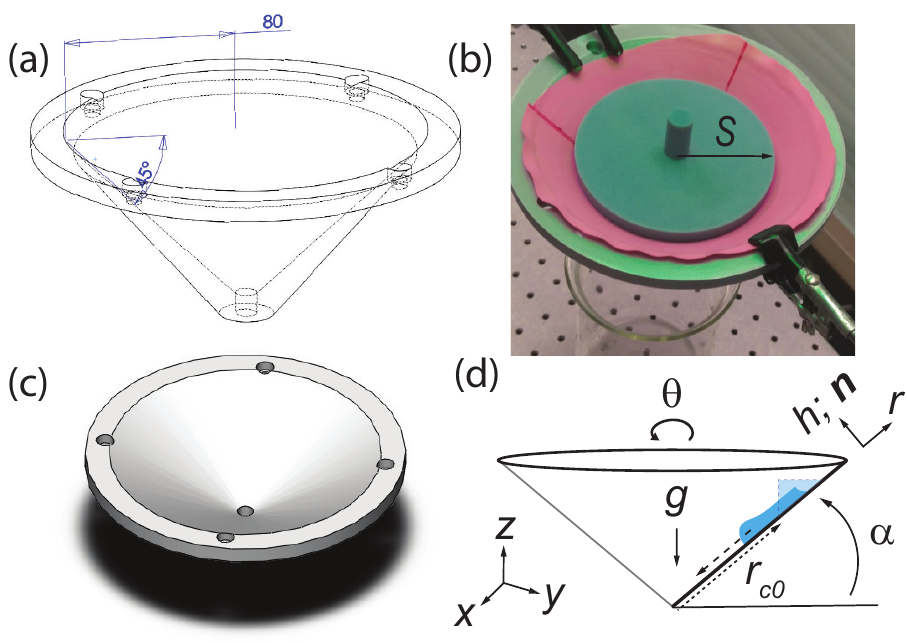}
\caption{(Color online) (a) Solidworks wire sketch of the 47$^o$ funnel.
Note that the technical drawing mentions 45$^o$; 
printing imperfections can somewhat modify the opening angle, which is therefore always measured after 
printing. (b) Photo of the experimental setup including the latex sheet, stopper and green light illumination. The beaker that collects the PDMS is visible underneath the funnel. $S$ indicates the radius of the stopper used as initial barrier that releases the fluid. (c) Solidworks drawing of the 35$^o$ funnel. (d) Coordinate system variables used in the description of the fluid behavior (solid lines). The light blue surface delineated with the dotted line indicates the initial fluid volume and position as created by the stopper at distance $r_{c0}$ (dotted arrow) from the origin of the funnel geometry; 
the thickness of the film referenced in the text, $h_i$, is measured in the vertical direction
(therefore, along the short stopper dimension). 
The dashed arrow indicates the flow direction of the thin film. The arrow denoted by $g$ indicates the direction of gravity.}
\label{fig:experiment}
\end{figure}

In the present paper, we focus on the funnel geometry, see figure~\ref{fig:experiment}, where initially a 
fixed amount of wetting fluid is deposited around a perimeter and then let to evolve due to gravity.  
Despite its relevance to a number of practical applications, funnel flow to the best our knowledge has not 
been yet carefully analyzed, in particular in the context of front instabilities. Funnel flow 
involves geometry-induced convergence, and the influence of this convergence, 
as well as of azimuthal curvature on instability development is unknown. 
For the purpose of understanding stability properties of a film in a funnel, it is useful to discuss some of the 
many limiting configurations that could be related to the considered one.  
If the film is deposited at a sufficient distance from the centre, the azimuthal curvature is small, and 
one could relate the problem to the finite volume of fluid deposited on an incline plane.  Even that problem is, 
however, difficult to analyze due to a time-dependent base state, see~\citet{hom91, Gomba2007}.  The limiting case of the opening angle 
$\alpha = 90^\circ$ could be thought of as a flow down a cylinder (in a direction of the cylinder axis)~\citep{Smolka2011,Mayo2013}, which shows fingering type of instabilities.  
Fingering instability is also observed for the flow down a surface of a cylinder
or a sphere~\citep{huppert_jfm_2010,balestra_jfm_2019}, the setup which shows similarity to fingering observed during spin coating~\citep{Melo,FH94}.   Another setup of interest is flow in a Hele-Shaw geometry where surface geometry plays a 
role in instability development, see~\citet{miranda_pre_2000,miranda_pre_2014}.  
In the limit $\alpha = 0^\circ$, 
one could think of the problem of closing a hole in a film on a horizontal substrate~\citep{DGG}, which is 
stable~\citep{2014_SM_capillary_collapse, bostwick_prf_2017,zheng_jfm_2018,hardt_jfm_2018}.  Another 
possibly relevant limit is that of a liquid filament, which on a horizontal substrate becomes unstable by a mechanism 
that could be related to Rayleigh-Plateau instability of a liquid jet, modified by the presence of a substrate~\citep{davis_jfm80}. 
That problem is, however, difficult to analyze in the limit of zero contact angle that we consider here, see~\citet{dgk_pof09}. 
Perhaps closer analogy is fluid ring on a horizontal substrate, which indeed may become unstable~\citep{gdk_jfm13}.  However,
the fact that there is no body force inducing converging flow makes this setup significantly different from the funnel flow.  
Converging nature of the flow in a funnel leads to thickening of the film, 
and since the film thickness is important in determining both the speed of spreading, and the instability mechanism itself,  
it is expected to influence the instability considerably. 
We should also point out that the problem opposite to our setting, rising film in a glass, was considered recently~\citep{DJFB20}.

Various limiting cases suggest many possible routes for analysis of the instability 
evolution.  In the present paper, we start by discussing our experimental results in Sec.~\ref{sec:exp}, and then follow 
in Sec.~\ref{sec:model} by considering appropriate
models for describing spreading of a film on curved substrates.
In Sec.~\ref{sec:results} we discuss first generic features of the funnel flow, by discussing similarities and 
differences to the flow down an incline, with particular focus on the regime such that useful input can be obtained by 
applying a self-similar approach.   Then, in Sec.~\ref{sec:exp_comparison} we apply the insights obtained in 
Sec.~\ref{sec:general} to interpret the experimental results, with particular focus on the instability development. 
Section~\ref{sec:conclusions} concludes the main part of the paper.  The LSA for a liquid film of constant flux flowing 
down an incline is briefly discussed in the Appendix.  Supplementary materials~\citep{SM} provide some technical details, experimental videos as well as the complete list of experimental results.

\section{The experiment\label{sec:exp}}

We designed funnels with Solidworks and 3D printed them on a fused-deposition 3D printer. Figure~\ref{fig:experiment} shows the details of the funnels. Inside the funnel, we glue a thin latex sheet, which helps create a smoother surface.
The funnel is placed between a green light source and a beaker that collects the liquid flowing out of the funnel opening. To prepare the experiment, a 3D printed stopper of radius $S = 6$~cm is inserted in the funnel, see figure~\ref{fig:experiment}(b). A known volume of fluid, $V$, is then carefully 
deposited around the funnel using a syringe; the fluid spreads itself evenly in the cavity around the stopper. This creates an initial film of height $h_i$ (measured in the vertical direction), 
which we use as our control parameter to simplify interpretation of the results (note that $V \approx \pi S h_i^2/\tan \alpha$). 
For every trial, the funnel and stopper are cleaned and leveled before deposition. 
For small values of the opening angle, $\alpha$, one needs a prohibitively large $V$ to keep $h_i$ in a range that is also appropriate 
for larger $\alpha$, so for such angles we choose smaller values of $h_i$.  
To start the flow, the stopper is raised and flow is observed. A fluorescent dye (Pyrromethene) is added to the fluid stock solution, 
enhancing the contrast between the moving fluid and the latex sheet under influence of the green light. 
We use polydimethylsiloxane (PDMS) of density $\rho$ = \SI{9.6e2}{\kilo\gram\per\meter^3}, 
viscosity $\mu$ = \SI{1e3}{\milli\pascal\second}, and 
surface tension $\gamma$ = \SI{21}{\milli\newton\per\meter}; for more details regarding 
PDMS properties see \cite{dijksman2019}.
Note that PDMS wet latex, since the critical surface-vapor surface tension of typical latex types $\approx 50$~mJ/m$^2$ is much higher 
than the low surface tension of PDMS $\approx 20$~mJ/m$^2$; see~\cite{2000_Ho_latex} and~\cite{Zhang_2018}. This means that the spreading parameter is larger than 0 and that thus the contact angle is zero, without hysteresis (see e.g. the book from~\cite{degennes2004capillarity}).
Elastocapillary effects such as discussed in ~\cite{2012_prl_elastocap} are neglected as the ratio of liquid-vapor surface tension to elastic 
modulus of the latex rubber is much smaller than the thickness of the latex sheet $\approx 1$~mm. Fluid flow is recorded using a high resolution camera at 
25 frames per second. The videos 
serve both to extract the wavelength of the developing instability and for the quantitative assessment of the flow speeds of the relevant film features. 

It should be pointed out that there are few experimental issues that lead to some variation in the extracted 
experimental results discussed later in the text (and denoted by error bars where appropriate).  At first, the method to 
distribute the fluid, while simple, may not have always led to a perfect azimuthally symmetric distribution. Another source of 
error is the formation of air pockets under the thin rubber sheet lining the funnels. Re-gluing prior to conducting experiments 
helped to create a surface free of larger surface abnormalities. Conducting multiple experiments and averaging the values helped to 
remedy some of these errors and reduce the error bars. More detailed information regarding the experiments, including selected 
experimental videos ({\it Videos 1 - 4}) as well as funnel specifications ({\it Drawing1, Drawing2}) are available~\citep{SM}; additional 
videos can be found at the NJIT Capstone Laboratory web page~\citep{web_capstone}. 

\begin{figure}
\centering
\includegraphics[scale=0.6]{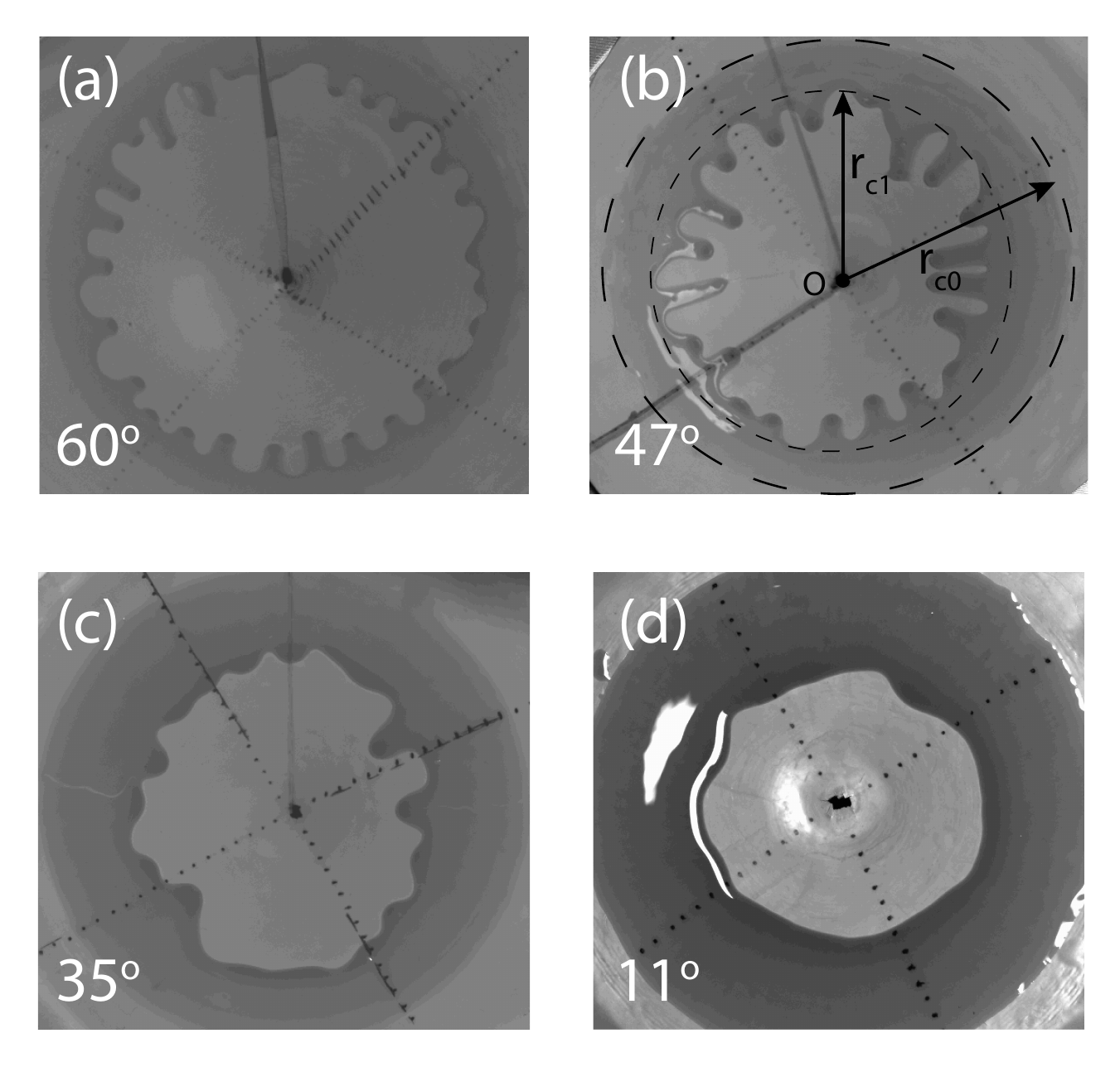}
\caption{(a-d) Flow examples for the 60, 47, 35 and 11 degree opening angles, respectively. The width of each image 
is  $\approx$ 60 mm, as indicated by the dark circle that demarcates the edge of the fluid volume before onset of the flow. 
See also (a) {\it Video1}; (b) {\it Video2}; (c) {\it Video3}; (d) {\it Video4}.}
\label{fig:experiment_data}
\end{figure}

\subsection{Extracting instability features}
The videos allow us to extract two main features of the instability: the number of fingers observed, $N_{\rm exp}$, and the onset 
radius of fingering, $r_{c1}$. We show snapshots from top view videos for four values of $\alpha$ to identify such features in 
figure~\ref{fig:experiment_data}(a-d). The experiments for each set of parameters are repeated several times to obtain conclusive results, which 
are summarized in Table~\ref{table:exp} that shows $N_{\rm exp}$, as observed for a few different values of $\alpha$ and $h_i$. We have also carried 
out additional experiments using PDMS of lower viscosity, $\mu$ = \SI{3.5e2}{\milli\pascal\second}, which show that $N_{\rm exp}$ is viscosity-independent, 
the point to which we return in Sec.~\ref{sec:exp_comparison}.  Before closing 
this section, we note that the rear contact line remains essentially fixed, 
with the fluid thinning in its vicinity, as it can be seen clearly in the 
supplementary videos~\citep{SM}.   This point will become relevant later when 
considering the self-similar solution.  

\begin{table}
\centering
\begin{tabular}{p{2cm}|p{2cm}|p{2cm}|p{1.75cm}|p{1.75cm}|p{1.75cm}}
\hline
 $\alpha$ & $V$ & $r_{c0}$ & $r_{c1}$ & $h_i$ & $N_{\rm exp}$\\
 (degrees) & (ml) & (mm)& (mm) & (mm) & (-)\\
 \hline
\multirow{2}{*}{60} & 2.8  &120  & 110 $\pm 0.7$ & 5   & 27 $\pm 1.2$\\
 & 5.4  &120  & 112 & 7 & 23\\
 \hline
\multirow{2}{*}{47} & 4.5  & 88 & 77 $\pm 1.2 $ & 5 & 21 $\pm 0.8$\\
 & 8.9  &88 & 75 $\pm 2.3 $ & 7 & 16 $\pm 0.5$\\
 \hline
35 & 6.9   &73   & 57 & 5 & 16 \\
 \hline
\multirow{2}{*}{11} & 6.7  &61 & 47 $\pm 3$ & 2.5 & 10 $\pm 2.8$\\
 & 9.7   &61   & 50 & 3 & 10\\
 & 17.8   &61  & 43 & 4 & 6\\
 \hline
\end{tabular}
\caption{Experimental results for the flow in a funnel as the opening angle and the initial
film thickness, $h_i$ are varied (the latter is controlled via varying fluid volume, $V$).   
The columns are the funnel angle, $\alpha$, the initial fluid volume, $V$, 
the initial distance to the funnel centre, $r_{c0}$, 
the distance at which instability is observed, $r_{c1}$, 
the initial film height (in the $z$ direction), $h_i$,
and the number of fingers observed, $N_{\rm exp}$. 
Errors reported are standard deviations. If no standard deviation is 
reported, only one video is available with a camera angle such that 
$r_{c1}$ could be extracted. 
Full data sets for experiments carried out also using 
PDMS of different viscosity, for additional fluid volumes 
and recorded using different camera angles are available, see {\it Supplementary Table1}.\label{table:exp}
}
\end{table}

\begin{figure}
\centering
\includegraphics[scale=0.5]{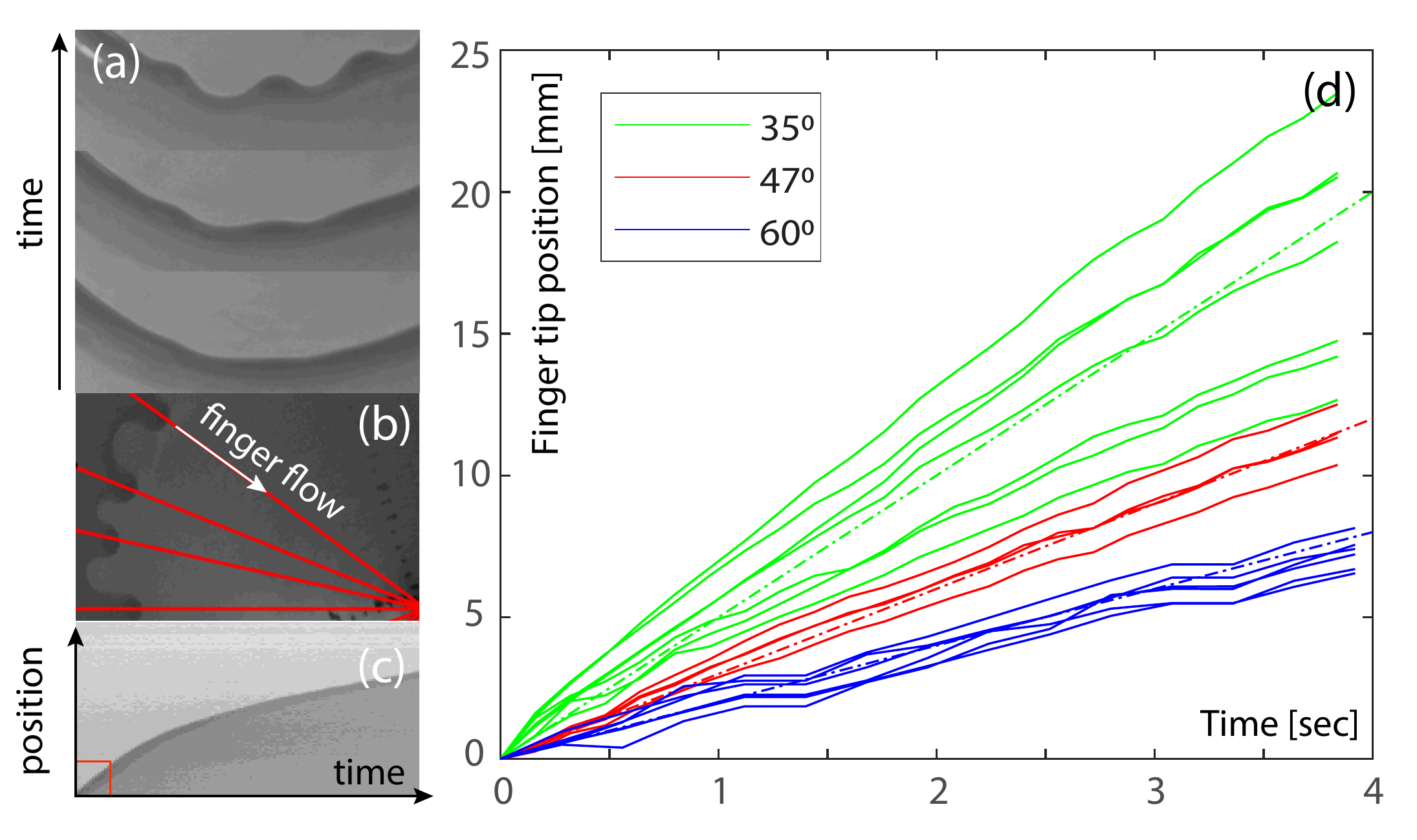}
\caption{(a) 
Front propagation as a function of time for the setting from figure~\ref{fig:experiment_data}. The middle part shows the onset of fingers, where they become countable and at which point we define $r_{c1}$. (b) The dark front edge allows for finger tip position tracking along lines (red) of pixels towards the centre of funnel (orifice). (c) A kymograph for the 60~degree, $h_i = 5$ mm shows the tip position as a function of time. The initial part in the red square (duration of which is approximately 4 seconds) features an initially constant tip displacement rate. Examples of this initial finger tip position dynamics are shown in (d) for three different opening angles $60^\circ,~47^\circ,~35^\circ$ and for the same initial $h_i$ = 5 mm. Dash-dotted line are straight lines as a guide to the eye to show the average tip speed for each value of $\alpha$.  
The time window shown in (d) corresponds approximately to the horizontal dimension of the red square in the kymograph shown in the part (c).  
Note an {\it increase} 
of the average speed with a {\it decrease} of $\alpha$, the point discussed further in Sec.~\ref{sec:exp_comparison}.}
\label{fig:experiment_data_fingers}
\end{figure}

\subsection{Extracting front speed}

To extract flow speed we need to extract quantitative data from the videos. In particular, we extract the finger 
tip position as a function of time. First, we need to know where $r_{c1}$ is, the onset position of fingering. 
Figure~\ref{fig:experiment_data_fingers}(a) illustrates in more detail our approach to finding this value. 
We define $r_{c1}$ by requiring that at the onset of fingering, the distinct undulations are present along the 
entire perimeter of the fluid front. We then need to extract the finger tip position for each finger. To that end, 
we define lines along the direction of motion of the fingers towards the funnel orifice.  Figure~\ref{fig:experiment_data_fingers}(b) 
shows a few of these lines as an example. The lines are shown in red; they all converge in 
the funnel orifice. Along these lines, we extract pixel values from the frames in the videos, which for every frame yield an intensity profile 
as a function of $r$. Due to the dark front of the finger, every intensity profile has a clear step which can be fitted with an 
error function to obtain a finger tip position as a function of time. The intensity profiles extracted for every finger and 
every frame can be put together to form a kymograph, indicating qualitatively the finger tip dynamics for the tip considered. 
An example of such a kymograph is shown in figure~\ref{fig:experiment_data_fingers}(c). Note the linear part in the first 
few seconds of the kymograph, and the nonlinear slowing down for later times. 
Figure~\ref{fig:experiment_data_fingers}(d) zooms into the early times and confirms that the finger
 tip velocities are constant, and also that they depend on the opening angle, $\alpha$.  
 This figure shows few examples of finger tip position as a function of time for three different 
values of $\alpha$. Tip positions are as measured from the finger starting point, which is always close to, but not exactly $r_{c1}$.  
Counterintuitively, the initial finger tip speed is larger for smaller $\alpha$'s, another point to which we return in 
Sec.~\ref{sec:exp_comparison}.

\section{The model}
\label{sec:model}

In this section we discuss the appropriate model for a liquid film in a funnel. 
Consider a funnel of opening angle $\alpha$, parametrized by
\[
(x, y, z) = (r\cos\alpha\cos\theta, r\cos\alpha\sin\theta, r\sin\alpha), \quad r\in[R_l, R_r], \quad \theta\in[0, 2\pi],
\]
where $0<R_l<R_r$. We can then define the orthogonal unit vectors on the funnel as
\[
\vect{e}_1 = (\cos\alpha\cos\theta, \cos\alpha\sin\theta, \sin\alpha), \vect{e}_2 = (-\sin\theta, \cos\theta, 0), \vect{n} = (-\sin\alpha\cos\theta, -\sin\alpha\sin\theta, \cos\alpha),
\]
where $\vect{n}$ is the unit normal vector pointing inside the funnel, see figure~\ref{fig:experiment}(d). The principle normal curvatures in the directions parameterized by $r$ and $\theta$ are given by
$
\kappa_1 = 0, \kappa_2 = \tan\alpha/r.
$
Based on \cite{Roy1997}, the evolution of the thickness of a thin liquid film, $h$, inside a funnel can be described by the following partial differential equation
\begin{eqnarray}
\left(1 - \frac{\tan\alpha}{r}\, h\right)\, h_t &=& -\frac{\gamma}{3\mu}\nabla_s\cdot\left[\left(h^3 - \frac{\tan\alpha}{2r}\,h^4 \right)\nabla_s \left(\nabla_s^2 h + \frac{\tan\alpha}{r - \tan\alpha h}\right) + \frac{\tan^2\alpha \, h^4}{2r^3}\,\vect{e}_1\right] \nonumber\\
& &  -\frac{\rho g}{3\mu}\nabla_s\cdot\left[-\sin\alpha h^3\left(1 -  \frac{\tan\alpha}{r} h\right)\,\vect{e}_1 - \cos\alpha h^3\nabla_s h \right],
\label{eq_full_dim}
\end{eqnarray}
where surface gradient, divergence and Laplace operators are defined by
\begin{eqnarray}
\nabla_s f &=& \frac{\partial f}{\partial r} \vect{e}_1 + \frac{1}{r\cos\alpha} \frac{\partial f}{\partial \theta} \nonumber
\vect{e}_2, \\
\nabla_s \cdot (q_1 \vect{e}_1 + q_2 \vect{e}_2)& = &\frac{1}{r}\, \frac{\partial}{\partial r} (r q_1) + \frac{1}{r\cos\alpha}\,\frac{\partial}{\partial \theta}(q_2), \nonumber  \\
\nabla_s^2 f &=& \frac{1}{r}\frac{\partial (r f_r)}{\partial r} + \frac{1}{r^2\cos^2\alpha} \frac{\partial^2 f}{\partial \theta^2}, \nonumber
\end{eqnarray}
respectively.  We nondimensionalize the problem by
$
h = a\, \bar{h}, r = a\, \bar{r},  t = t_c \, \bar{t}, 
$
where $a=\sqrt{\gamma/\rho g}$ is the capillary length, and $t_c = 3\mu a/ \gamma$ is the timescale. 
\citet{Howell2003} pointed out that for a thin film such that $h \ll r/\tan(\alpha)$, 
the model can be simplified by neglecting asymptotically small terms; 
after dropping the bars, the governing equation is 
given by 
\begin{equation}
\frac{\partial h}{\partial t}  = -\nabla_s\cdot\left\{h^3\left[\,\nabla_s \left(
\underbrace{\nabla_s^2 h}_\text{surface tension} + 
\underbrace{\frac{\tan\alpha}{r}}_\text{substrate curvature}
\right) 
-\underbrace{\sin\alpha \vect{e}_1}_\text{tangential gravity} 
-\underbrace{\cos\alpha \nabla_s h}_\text{normal gravity}
\right]\right\}.
\label{eq:short}
\end{equation}
For the experimental parameters given in Sec.~\ref{sec:exp}, we have $a \approx 0.15$ cm and
$t_c \approx 0.2$ s. For the consistency with the experiment, we choose the computational domain $r\in[1, L]$, $L=100$, 
$\theta\in[0, 2\pi]$ and  $h= O(1)$.  The $r=1$ is chosen as the domain boundary so to avoid the
coordinate singularity at $r=0$.

The computational results that we discuss in Sec.~\ref{sec:results} are obtained by implementing
second-order Crank-Nicolson method in time, second-order discretization in space and Newton's method to 
solve the nonlinear system at each time step, as described in detail in, e.g.,~\cite{Lin2010}.
To deal with the well-known issue of contact-line singularity, it is appropriate to introduce matched asymptotic expansions to join solutions in different length scales near the contact line, see \cite{Snoeijer2013, Sibley2015} for further details. 
One can also introduce the interface potential that in general gives rise to an equilibrium film thickness. This film plays the role of a microscopic length scale that, again, has to be matched with outer solution where viscous forces are not important, see \cite{Pismen2008}. 
We, however, for simplicity 
assume directly that the 
solid substrate is prewetted, i.e., already covered by a thin layer of fluid.  Assuming the 
presence of such a prewetted layer essentially removes the contact line from the consideration.
While various other models including relaxation of no-slip boundary condition exist and could 
be implemented, it is known that from the macro-scale point of view (that is, consideration of the
film dynamics) what really matters is the lengthscale that is introduced by a model~\citep{DKB01}.
In particular, for a simple and well researched
constant flux flow down an incline (where time-independent influx leading to a fixed film thickness
far behind the front is assumed), it is known that there is a translationally invariant
solution for a film moving down an incline with the speed $U$ that only weakly depends on  
the precursor film thickness, $b_0$, as long as $b_0 \ll 1$~\citep{BB97}.  
It should be noted though that the limit $b_0\rightarrow 0$ is singular, leading to a shock-like
singularity; the details of the film behavior in the presence of a vanishingly small length scale
have been considered extensively in the literature, see, e.g.~\cite{cm_rmp09,bonn_rmp2009}; we do not discuss them further in the present work.   

We note that while for a flow down an incline it can be 
assumed that the precursor film thickness is a constant (independent of position), for the flow 
in a funnel, conservation of fluid volume requires that the flux at inlet and outlet are the same.  
One simple choice of the boundary conditions that satisfies this condition is 
\begin{equation}
h(r=1) = b_0\left(\frac{\frac{1}{L} + L\cos\alpha}{1 + \cos\alpha}\right)^{1/3},\,
h(L) = b_0, \,
\left.\nabla_r\left(\nabla_r^2 h\right) - \cos\alpha \nabla_r h\right|_{r=1, \,L} = 0.
\label{eq:bc}
\end{equation}
The precursor film thickness $b(r)$ is obtained as the time-independent solution 
of one dimensional version of Eq.~(\ref{eq:short}) (where the solution is assumed to be $\theta$-independent), 
and with the boundary conditions as specified by Eq.~(\ref{eq:bc}).
The solution of this nonlinear boundary value problem is found using Matlab's ``fsolve''.

\section{Results}
\label{sec:results}

In this section we present the results of analysis, simulations and comparison between theoretical predictions with experiments. 
In Sec.~\ref{sec:general} we focus on understanding the influence of funnel geometry on the flow
without immediately attempting to develop direct comparison with the experiments.  In this section we also 
discuss the insight that could be reached based on application of a self-similar type of approach.  
Then, in Sec.~\ref{sec:exp_comparison} we focus on the comparison of the theoretical and the experimental results. As we will see, an useful insight can be reached by developing a connection between the funnel 
flow and flow down an incline plane.

\subsection{Film flow in a funnel: General considerations}
\label{sec:general}

\begin{figure}
\centering
\includegraphics[scale=0.6]{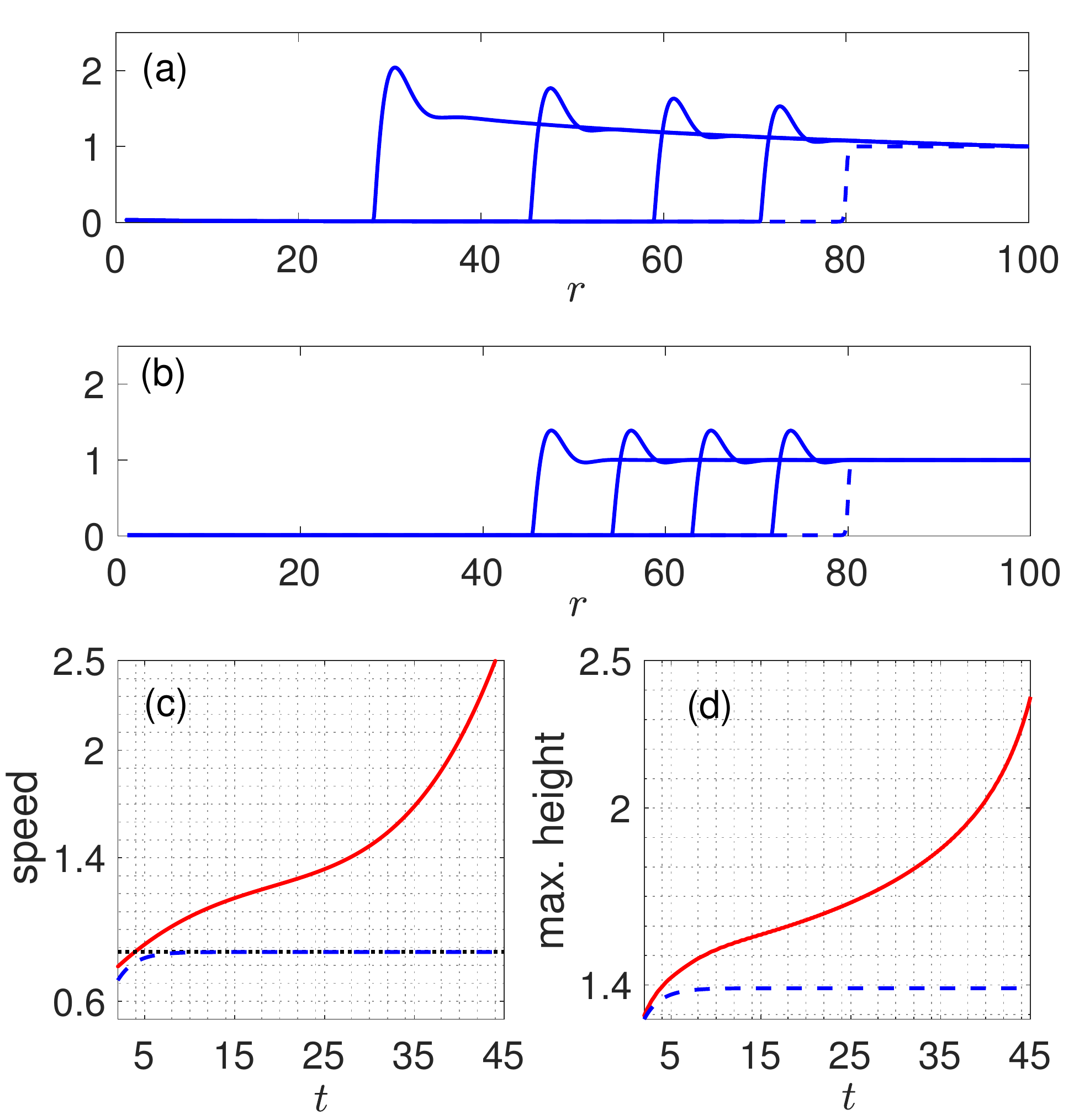}
\caption{
(Color online) Constant flux flow.  
(a)  Funnel: azimuthally symmetric flow; (b) Incline: unperturbed flow (for simplicity 
we use the same variable $r$ for the both flows; in (b) $r$ stands for the downhill coordinate). 
The initial condition (dashed) for both (a) and (b) is specified by Eq.~(\ref{eq:ic_cf}). 
The film profiles are shown at times $10$, $20$, $30$ and $40$ (solid lines). 
Panels (c-d) shows the speed (c) and the maximum height of the film (d) for the funnel flow (solid red) and 
for incline plane flow (dashed blue). 
The dotted (black) line in (c) shows the flow speed 
for a flow down an incline as discussed in the text; the difference between the computed and theoretical 
speed illustrates the (minor) influence of the initial 
transient behavior.  Here, the inclination angle is $\alpha = 60^\circ$, the initial front position is 
$r_{c0} = 80$ and $b_0 = 0.01$.  
}
\label{fig:cf}
\end{figure}

\subsubsection{Constant flux flow} 

For the incline plane flow, the best known case is the constant flux configuration and so 
we start by considering such a setup in a funnel.
The initial film profile specified at $t=0$ is a (smoothed) rectangular profile of the unit height as follows
\begin{equation}
h(r, t=0) = b(r) + \frac{1-b_0}{2}\left(1+\tanh(5(r-r_{c0}))\right),
\label{eq:ic_cf}
\end{equation}
where $r_{c0}$ corresponds to the front position.
Figure~\ref{fig:cf}(a) shows the profiles that develop at different times.  
To better illustrate the influence of funnel geometry on the flow, we also show in (b) the results for 
constant flux flow down an incline. The latter results are obtained by solving numerically 
Eq.~(\ref{eq_x}) (see Appendix~\ref{app:incline}), similar to the ones presented in e.g.~\cite{Lin2010}, 
with a uniform precursor film, $b(r)\equiv b_0$, and consistent boundary conditions. 
Both sets of simulations show the formation of a capillary ridge behind the front, as expected.  
The comparison of the results for the flow in a funnel and down an incline shows that for the former the 
film thickness is generally larger due to converging flow nature.  Since the speed of the front is expected to scale 
with the film height as $U \propto h^2$~\citep{Hupp82}, 
this thickening also leads to a faster flow down a funnel compared to the flow down an incline, see 
figure~\ref{fig:cf}(c).  After initial transients, the latter 
evolves to a travelling wave moving with a constant speed, see figure~\ref{fig:cf} (c-d).   Within the
presently used scaling, this speed is given by $U \sim h^2  \sin \alpha$, see Appendix~\ref{app:incline} and
note that rescaled quantities are used there.  The choice of relevant film thickness, $h$, entering this 
relation becomes more complicated for the constant volume flow, discussed in what follows.

\begin{figure}
\centering
\includegraphics[scale=0.6]{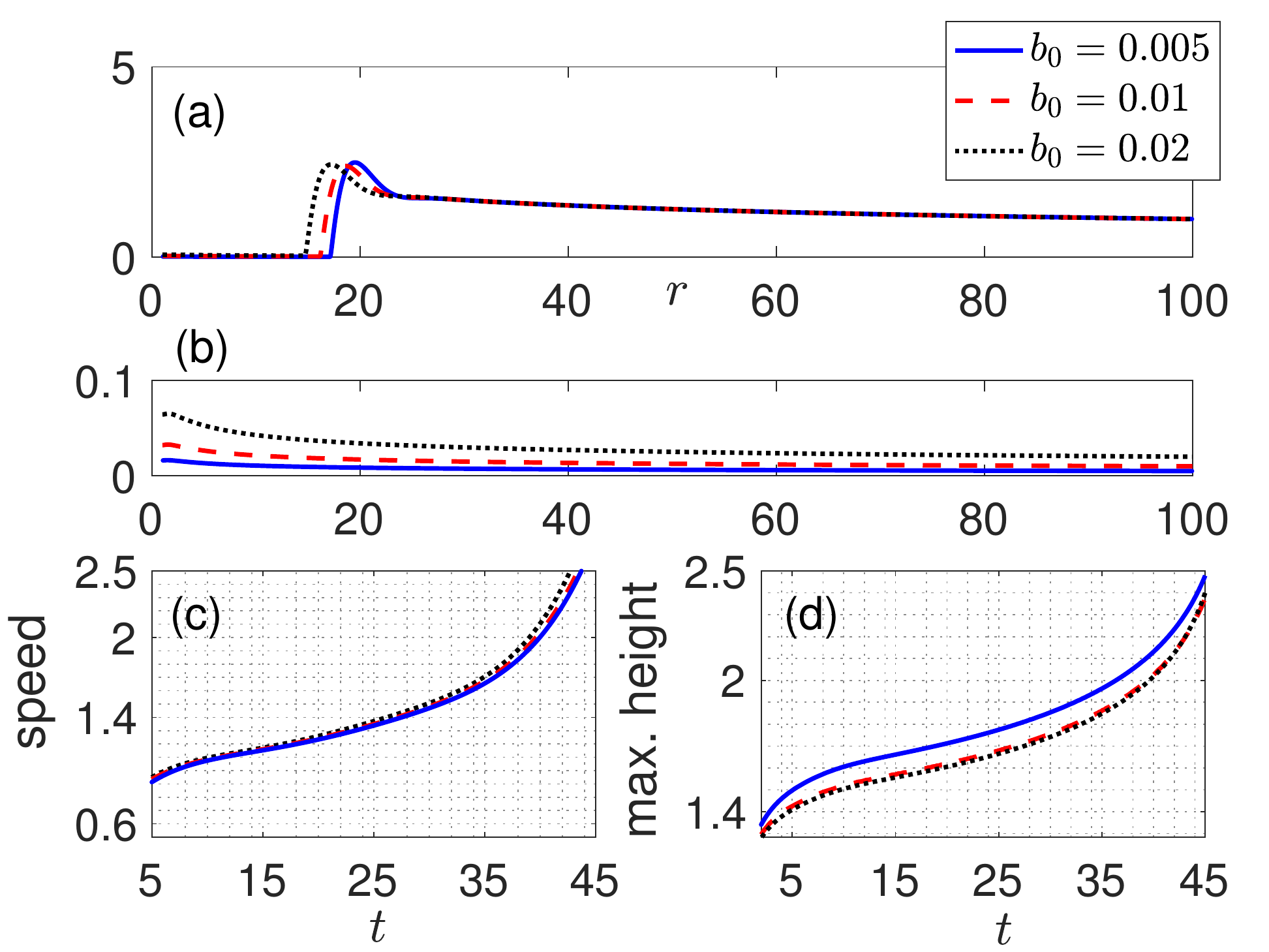}
\caption{(Color online) Constant flux flow in a funnel: 
influence of the  precursor film thickness for azimuthally 
symmetric flow. 
In all parts of the figure, the solid (blue), dashed (red) and dotted (black) lines corresponds to $b_0=0.005$, $0.01$ and $0.02$, respectively.
(a) Film profiles at $t=45$, (b) precursor films, (c) speed of the spreading front, and (d) maximum film height.  Here  $\alpha=60^{\circ}$ and $r_{c0}=80$.
}
\label{fig:precursor}
\end{figure}

Before proceeding with consideration of constant volume flow, we digress briefly to comment on the influence of precursor film thickness 
on the results. Figure~\ref{fig:precursor} shows an example of the results 
obtained for $b_0 = 0.005$, $0.01$ and $0.02$.  We recall that for a film flowing down an incline plane in
constant flux configuration, for larger precursor film thickness the spreading speed is (slightly) larger, see
Appendix~\ref{app:incline}.  
The same trend is found for the flow in a funnel, see figure~\ref{fig:precursor}(a, c).  Figure~\ref{fig:precursor}(b)
illustrates the steepening effect for the precursor film itself close to the funnel center, at small values of $r$.  

\subsubsection{Constant volume flow: Self-similar approach}
\label{sec:self-similar}

Next we study the spreading of constant volume film in a funnel, focusing first on the 
insight that can be reached by considering the regimes where a self-similar solution 
can be formulated.  For the flow down an incline, the self-similar  solution~\citep{Hupp82} (ignoring surface tension effects), predicts that the front speed scales as $t^{-2/3}$, and the height 
behind the ridge, $h_0$, as $t^{-1/3}$.   The question is whether a similar approach 
could be used for the funnel flow.  

Let us consider only the effect of substrate curvature and tangential gravity, and neglect all the other terms in Eq.~(\ref{eq:short}). 
This simplification (valid sufficiently far behind the film front and for the opening 
angles which are not too small) leads to
\begin{equation}
h_t = \frac{1}{r}\left[rh^3\left(\frac{\tan\alpha}{r^2}+\sin\alpha\right)\right]_r.
\label{eq:tangential}
\end{equation}
We observe that the substrate curvature amplifies the parallel component of gravity, 
and the amplification is larger when the flow is closer to the funnel centre. 
Next, we assume a solution of the form
\[
h(r, t) = T(t)H(\eta), \quad \eta =\frac{R-r}{r_f(t)}, 
\]
where $R$ specifies the position of the uphill part of the deposited fluid (assumed to 
be a constant, which is a good approximation of the experiment, as discussed 
in Sec.~\ref{sec:exp}), $r_f(t)$ is the distance travelled by the front 
and $R-r_f(t)$ is the position of the front, both at time $t$ (we drop the specific dependence on $t$ from now on).  
Equation~(\ref{eq:tangential}) then leads to
\begin{equation}
\frac{\dot{T} r_f}{T \dot{r}_f} H - \eta H' = \frac{T^2}{\dot{r}_f}
\left[
\frac{r_f}{R-r_f\eta}
\left(\sin\alpha-\frac{\tan\alpha}{(R-r_f\eta)^2}\right)
H^3 - 3\left(\sin\alpha+\frac{\tan\alpha}{(R-r_f\eta)^2}\right)H^2H'
\right],
\end{equation}
where the over-dot notation denotes the time derivative, and primes denote the derivative with respect to $\eta$.  
The solution should satisfy the volume conservation condition
\begin{equation}
2\pi\int^{R}_{R - r_f(t)} r h(r, t)\,dr = 2\pi T r_f\int^1_0 (R-r_f \eta)H(\eta)d\eta=v_c,
\label{eq:vol1}
\end{equation}
where $v_c$ is the fluid volume. 

At early times after fluid deposition,  $r_f\ll R$, and at the leading order in the small quantity $r_f/R$ we obtain
\begin{equation}
\frac{\dot{T} r_f}{T \dot{r}_f} H - \eta H' = -3c_s\frac{T^2}{\dot{r}_f}H^2H',
\label{eq:sim2}
\end{equation}
where $c_s=\sin\alpha+\tan\alpha/R^2$, 
and the volume constraint at the leading order reads
\begin{equation}
T r_f\int^1_0 H(\eta)d\eta=\tilde v_{c}, 
\label{eq:vol2}
\end{equation}
where $\tilde v_{{c}}=v_c/(2\pi R)$. 
We note that Eqs.~(\ref{eq:sim2}) and (\ref{eq:vol2}) are identical 
to those derived for the incline plane problem~\citep{Hupp82}.  
Simple scaling arguments give $T(t)\sim t^{-1/3}$, $r_f(t)\sim t^{1/3}$ and therefore the self-similar solution is
\begin{equation}
h(r, t) = \frac{1}{\sqrt{3c_s}}\sqrt{\frac{R-r}{t}}.
\label{eq:selfsimi}
\end{equation}
The volume conservation constraint, Eq.~(\ref{eq:vol2}), gives the location of the leading edge 
\begin{equation}
r_f(t) = \left(\frac{27c_s \tilde v_{c}^2}{4}\right)^{1/3} \,t^{1/3},
\label{eq:self_sim_rf}
\end{equation}
and the film height at the front 
\begin{equation}
h(r=R-r_f, t) = \left(\frac{\tilde v_{c}}{2c_s}\right)^{1/3} \,t^{-1/3}.
\label{eq:self_sim_h}
\end{equation}
This result shows that, in the limit when the fluid front only travels a short enough distance, 
the flow down a funnel is identical to the flow down an incline plane, the result which may 
not be immediately obvious.

\begin{figure}
\centering
\includegraphics[scale=0.6]{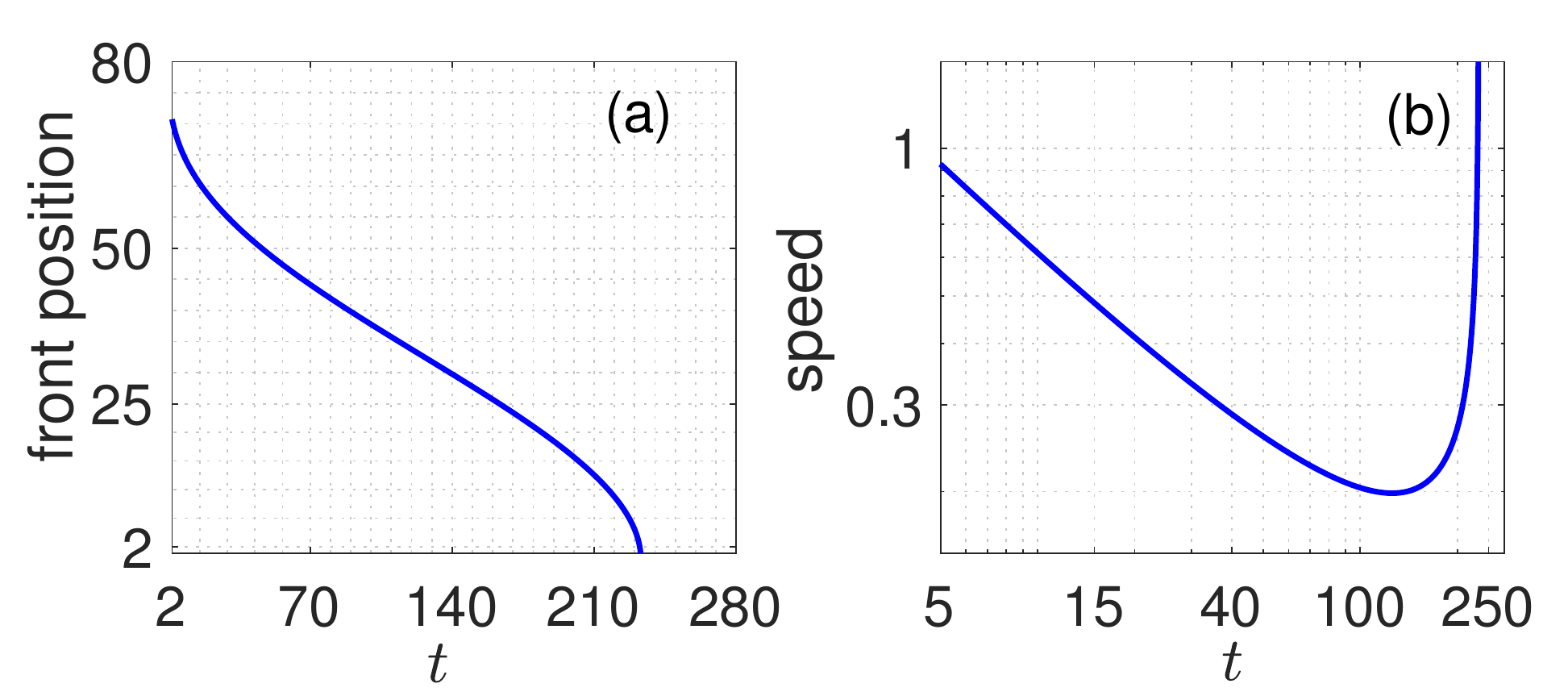}
\caption{(Color online) Constant volume flow in a funnel: predictions based on self-similar approach.
 (a) Front position ($R-r_f$), (b) front speed ($dr_f/dt$) in Eqs.~(\ref{eq:rf}) and (\ref{eq:drf}) (note log-log scale in (b)). Here $R=90$ and $\hat{c}=1370$, corresponding to the same funnel angle and volume used later in Fig.~\ref{fig:cv}. 
}
\label{fig:selfsimi}
\end{figure}

\subsubsection{Constant volume flow: convergence effects}
To gain some insight regarding the influence of the convergent nature of the funnel flow, we use the self-similar solution
specified by Eq.~(\ref{eq:selfsimi}) as an ansatz (but one should keep in mind that this solution is only valid for $r_f\ll R$) 
and require that the complete volume conservation constraint, Eq.~(\ref{eq:vol1}), should be satisfied. 
Following this approach, we find that $r_f$ satisfies the following equation
\begin{equation}
\frac{2R}{3}r_f^{3/2} - \frac{2}{5} r_f^{5/2} = \hat{c}\sqrt{t}, 
\label{eq:rf}
\end{equation}
where $\hat{c}=\sqrt{3c_s}v_c/2\pi$. 
By taking the time derivative, we obtain the equation for the front speed
\begin{equation}
\frac{d r_f}{dt} = \frac{\hat{c}}{2(R r_f^{1/2} - r_f^{3/2})\sqrt{t}}.
\label{eq:drf}
\end{equation}
Figure~\ref{fig:selfsimi} shows the solution for the front position 
($R-r_f$) and the front speed ($dr_f/dt$).  We see that the convergence effect leads to acceleration 
of the front for later times.  
To analyze this acceleration in more detail, we note that for the later stage of spreading, when the 
front is close to the funnel centre, we may assume $R - r_f\ll R$.  For the discussion of this regime, 
it is convenient to introduce the stopping time 
$T_s = (4/(15 \hat{c}))^2\, R^5$, at which the fluid front reaches the 
funnel centre, $r_f (T_s) = R$. We can then rewrite Eq.~(\ref{eq:rf}) as
\begin{equation}
\frac{2}{3}\left(1 - \frac{R-r_f}{R}\right)^{3/2} - \frac{2}{5} \left(1 - \frac{R-r_f}{R}\right)^{5/2} = \frac{4}{15}\sqrt{1 - \frac{T_s-t}{T_s}}.
\label{eq:rf2}
\end{equation}
At the leading order in the small quantity $(R-r_f)/R$, we obtain $R-r_f =\sqrt{\frac{4R^2}{15 T_s}} \sqrt{T_s - t}$. 
Therefore, the front speed $\dot{r}_f$ scales as $(T_s-t)^{-1/2}$ showing that the front is expected to accelerate 
when approaching the funnel centre.

\begin{figure}
\centering
\includegraphics[scale=0.55]{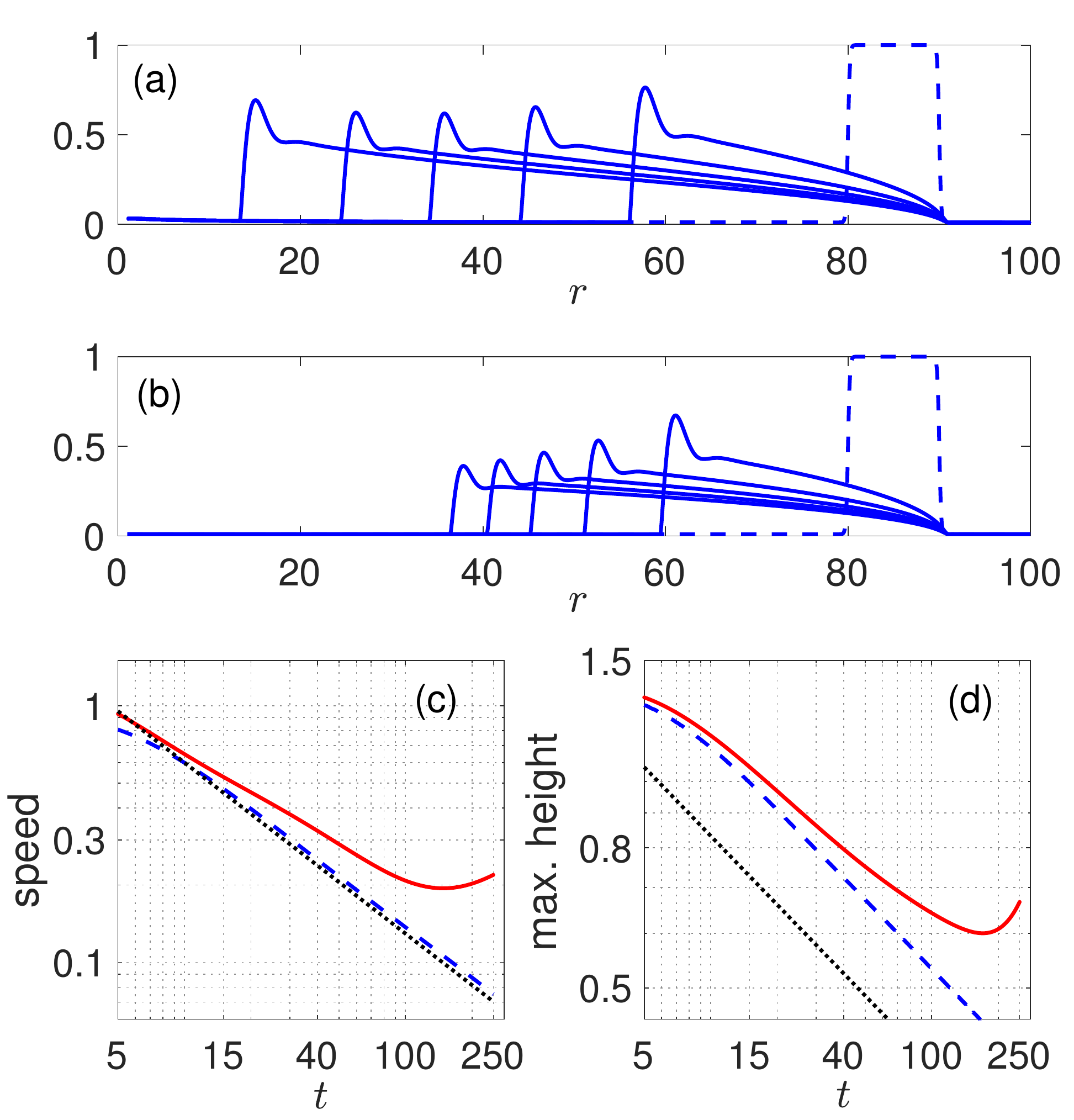}
\caption{ 
(Color online) Constant volume flow. 
(a)  Funnel: azimuthally symmetric flow; (b) Incline: unperturbed flow.  
The initial condition (dashed) is the same for (a~-~b) and is specified by Eq.~(\ref{eq:ic}). 
The film profiles are shown at times $50$, $100$, $150$, $200$, and $250$ (solid lines).
(c~-~d)  
The speed (c) and the maximum height of the film (d) for the funnel flow (solid red) and incline plane 
flow (dashed blue) (note log-log scale). 
The dotted (black) lines in (c) and (d) plot 
the self-similar solutions (Eq.~(\ref{eq:self_sim_rf}~-~\ref{eq:self_sim_h})), 
respectively; note that Eq.~(\ref{eq:self_sim_h}) applies to the thickness behind
the capillary ridge, leading to an approximately constant offset to the numerical solution for the 
maximum height. 
Slight deviation of the numerical solution from the 
expected scaling for very early times illustrate the minor influence 
of the initial transient behavior. 
Here, the inclination angle is $\alpha = 60^\circ$, the initial front position is 
$r_{c0} = 80$, $w=10$ and $b_0 = 0.01$.  
}
\label{fig:cv}
\end{figure}

\subsubsection{Constant volume flow: Numerical solution}
\label{sec:numerical}

Next, we study the spreading of a constant volume film in a funnel utilizing numerical simulations. 
The initial film profile at $t=0$ is specified by the following expression
\begin{equation}
h(r, t=0) = b(r) + \frac{1-b_0}{2}\left(\tanh(5(r-r_{c0})) + \tanh(5(r_{c0}+w-r))\right),
\label{eq:ic}
\end{equation}
where $r_{c0}$ corresponds to the front position, and $w$ determines the fluid volume. 
Figure~\ref{fig:cv} shows the film profiles, both for 
(a) funnel geometry and (b) for the same fluid volume travelling 
down an incline plane.   For the flow down an incline, we 
observe film thinning, as expected.  For the funnel flow, 
we observe different behaviour, with the thinning effect significantly
reduced, or even inverted for the later times.  As a consequence, the film in a funnel spreads 
significantly faster.

The parts (c) and (d) of this figure show 
that the scaling laws predicted by the self-similar solution~\citep{Hupp82} are
accurately reproduced for the constant volume flow down an incline.  Regarding the 
front speed shown in figure~\ref{fig:cv}(c) and ignoring transient effects for
very early times, the self-similar solution 
specified by Eq.~(\ref{eq:self_sim_rf}) captures precisely its behaviour for early 
times, including the prefactor.  
Regarding figure~\ref{fig:cv}(d), note that here we plot numerical result for 
the maximum film height, not the height behind the ridge to which the similarity 
solution applies; however, since the behaviour of the two considered quantities is 
essentially the same, the power law expected from the self-similar solution, Eq.~(\ref{eq:self_sim_h}), captures well the behavior of the maximum height for 
early times of the evolution, modulo a (constant) offset.

Focusing next on the funnel flow, we note the speed-up of the fluid front for 
the late times, as predicted by the self-similar solution derived in Sec.~\ref{sec:self-similar}.  
This speed-up is not as strong as predicted (viz. figure~\ref{fig:selfsimi}(b)), which is not surprising since the self-similar approach is not expected to be accurate for late times. 
Figure~\ref{fig:cv}(d) also shows the corresponding increase of the film height.
This thickening effect, which is also relevant for the intermediate times shown in 
figure~\ref{fig:cv}, is responsible for the deviation from the spreading law predicted by the self-similar 
solution (note deviation of the slope of the red lines in figure~\ref{fig:cv}, parts (c) and (d), from 
the scaling expected by the self-similar solution for the flow down an incline).

\begin{figure}
\centering
\includegraphics[scale=0.55]{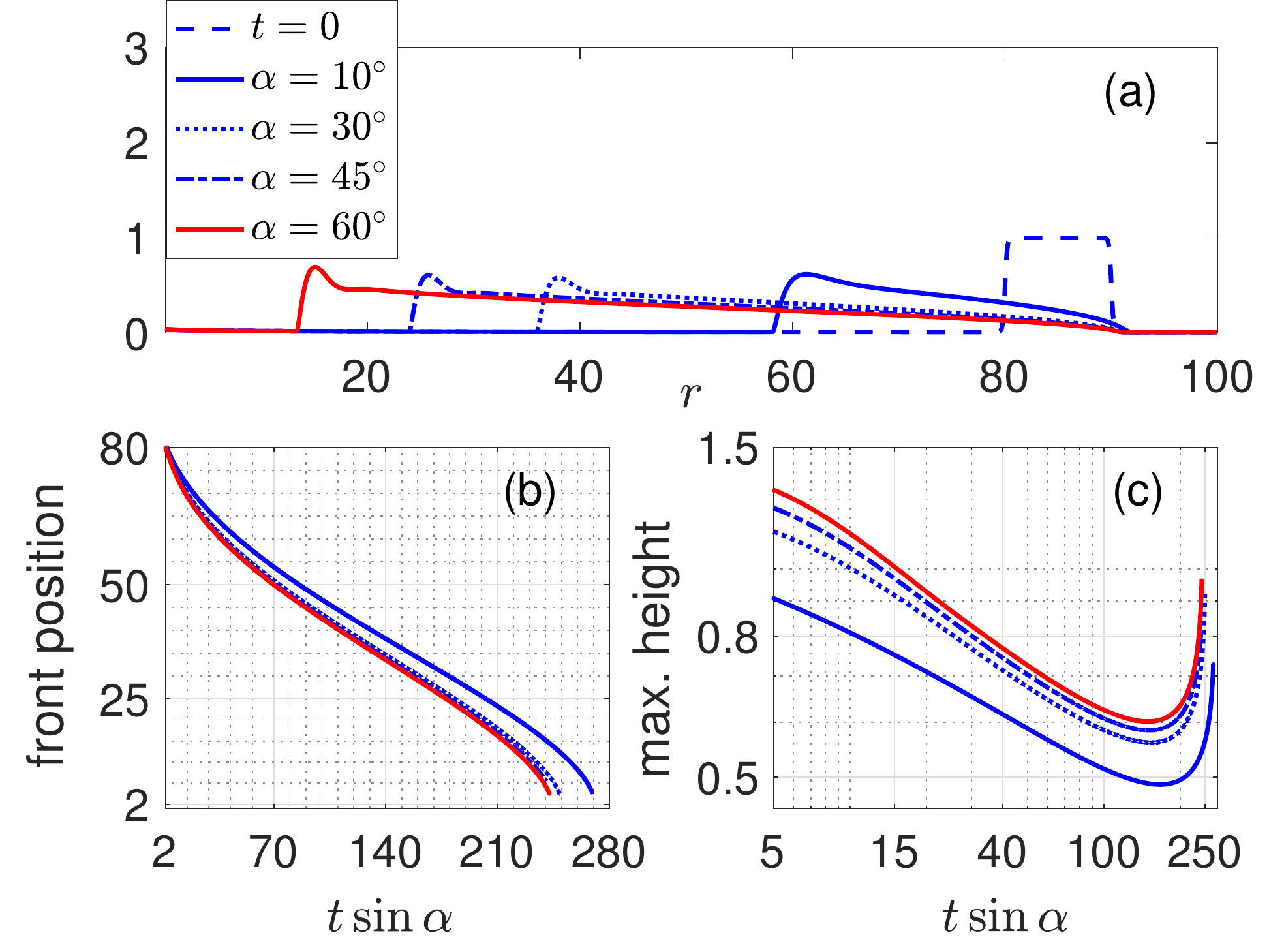}
\caption{(Color online) Funnel flow: Constant fluid volume spreading as the opening angle $\alpha$ is varied.  
(a) Film profiles at $t=250$. The initial conditions are taken to be the same, shown by the dashed line. 
The position of the spreading front and the maximum film height are shown in (b) and (c), respectively (note log-log scale in (c)). 
Here, the initial front position is $r_{c0} = 80$, $w=10$ and $b_0 = 0.01$. 
}
\label{fig:cv_alpha}
\end{figure}

Figure~\ref{fig:cv_alpha} shows the results obtained for the constant volume flow in the funnel geometry 
as the opening angle, $\alpha$, is varied.  The film spreads faster down a funnel characterized by 
larger $\alpha$, as shown in part (a).  We suspect that the tangential gravity 
may have dominant effect on the time scale of the flow; to show that this is the case, we plot the results
for the front position and maximum film height versus $t\sin\alpha$ in the parts (b) and (c).
We find approximate collapse of the front position curves in the part (b), showing that indeed 
the tangential gravity plays the major role.  Regarding the maximum film height shown in the part (c), 
we observe that this quantity is larger for larger $\alpha$'s, as expected since the capillary ridge
is more pronounced for such angles (we expect that this effect is also responsible for slightly 
faster spreading for larger $\alpha$'s observed in part (b)).  However, the trend of the maximum 
heights is similar for all $\alpha$'s, with the film height decreasing for early times, while at the later times 
when the front reaches closer to the funnel centre, the film height increases, see part (c), 
and faster spreading is observed, see part (b).

\subsubsection{Instability development}

To obtain a basic idea regarding instability development (finger formation), we discuss first the flow down an incline plane for 
the constant flux configuration.  In such a setup, the base  
state (for which the film thickness does not depend on the transverse coordinate), 
translates down an incline at a constant speed $U$, as already discussed.  
This fact allows for carrying out the linear stability analysis in the moving frame translating (with speed $U$) 
with the film itself; in this frame the base state is time independent~\citep{BB97}.
Appendix~\ref{app:incline} briefly outlines this problem, and discusses in particular the wavenumber of maximum growth, $q_m$, 
the corresponding wavelength, $\lambda_m = {2 \pi/q_m}$, as well as the critical 
wavenumber $q_c$, such that the wavenumbers $ q> q_c$ are stable; see figure~\ref{fig:LSA_D} in the Appendix~\ref{app:incline}.   
The stability analysis becomes more complicated for the constant volume flow down an incline, 
see~\citet{Gomba2007}, since for that problem the base state itself is evolving, as also illustrated 
in figure~\ref{fig:cv}(b). For the flow in a funnel, viz.~figure~\ref{fig:cv}(a), an 
additional complication involves gradual thickening of the film due to convergent flow.  

\begin{figure}
\centering
\includegraphics[scale=0.6]{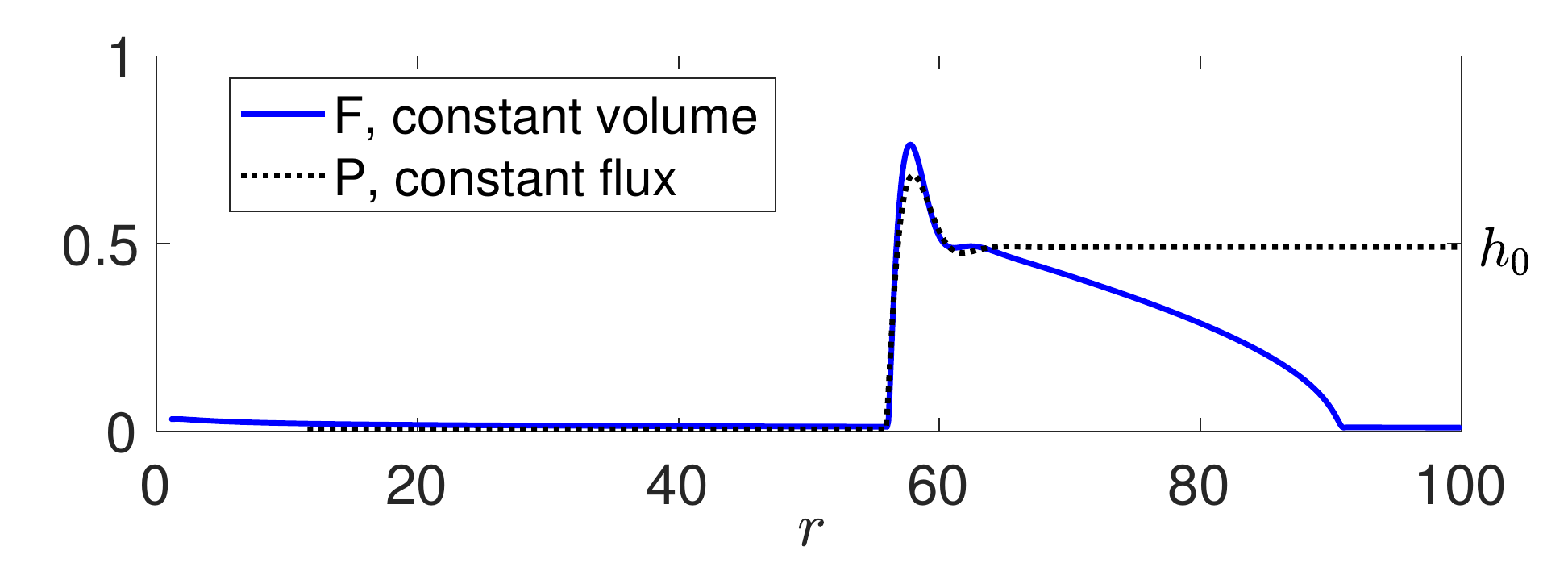}
\caption{(Color online) 
Film profiles in a funnel ('F', constant volume) and on an incline plane ('P', constant flux). 
The former one is the film profile at $t=50$ in figure~\ref{fig:cv}(a).
The later one is a travelling wave solution with the film 
thickness behind the front corresponding to the thickness at the inclination of point of `F'.
This thickness, $h_0$, determines the scale that is used in the LSA. 
}
\label{fig:LSA}
\end{figure}

In Sec.~\ref{sec:exp_comparison} we will consider rather simple approach to 
utilize the LSA results in the incline plane problem to compare with experiment; 
here we outline the basic aspect of this approach, 
without explicit reference to the experiment.  
Let us consider constant volume flow in a funnel, as shown in figure~\ref{fig:LSA}.  
When the film front has reached a prescribed position for the considered opening 
angle, $\alpha$, the thickness $h_0$ is extracted from the film profile as the 
thickness at the inflection point behind the capillary ridge. 
With the knowledge of this characteristic thickness, $h_0$, and the opening angle, $\alpha$, we can 
then find a travelling wave solution on an incline plane that has exact the same characteristic thickness; 
such solution is plotted in figure~\ref{fig:LSA} as well (marked by P).  
The LSA results of this travelling wave solution then gives us the most unstable wavenumber. 
Therefore, this most unstable wavenumber results from combination of the information from experiments (instability location), numerical simulations of funnel flow (providing $h_0$), and the LSA 
originating from the flow down an inline plane.
In the next section, we discuss how to use similar approach to provide basic understanding of the instability development in 
experiment.  

\subsection{Film flow in a funnel: Comparison with the experiment}
\label{sec:exp_comparison}

\begin{figure}
\centering
\includegraphics[scale=0.6]{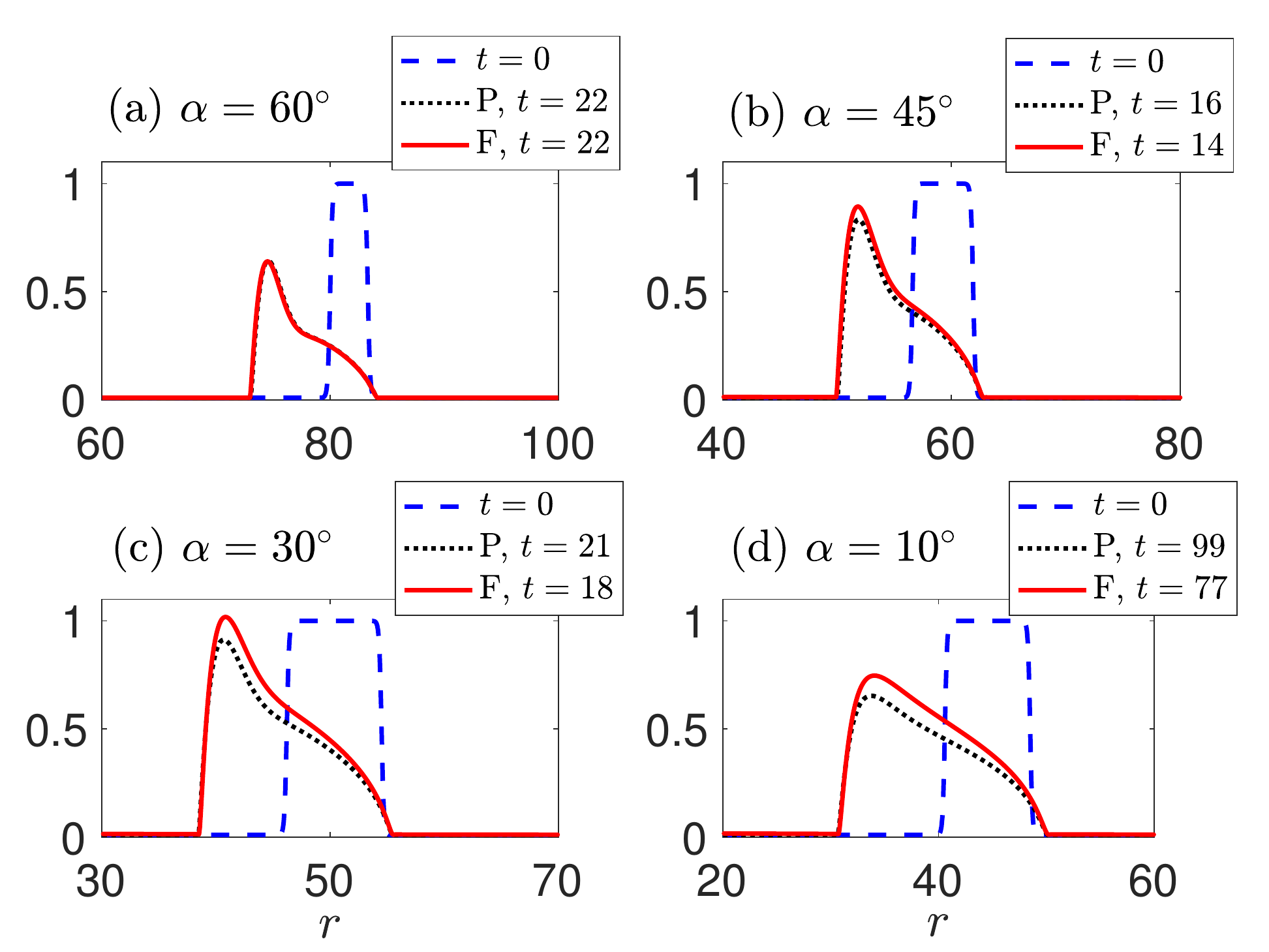}
\caption{(Color online) Time evolution of a film on an incline plane ('P') and in a  funnel ('F'). 
The initial condition is specified by Eq.~(\ref{eq:ic}). 
The results are plotted at the times at which experimentally 
observed instability radius, $r_{c1}$, is reached. Note that the shown range of $r$, $40$, is the same for all figure parts.
The initial volume correspond to the experimentally used one for the film 
height of $5$ mm for $\alpha = 60^\circ,~47^\circ,~35^\circ$, and of $2.5$ mm for $\alpha = 11^\circ$, see
Table~\ref{table:exp}.  
}
\label{fig:funnel_cv_exp}
\end{figure}

We now proceed to the consideration of a funnel flow, but with the specific 
emphasis on the comparison with the experiments.  While we will modify the 
choice of the parameters that we use to more closely resemble the 
experimentally relevant ones, for simplicity we still keep the smoothed
rectangular initial profile, with the idea that the instability takes 
some time to develop, and therefore the initial film profile is not of relevance.
However, we do choose the initial film width, $w$, 
see Eq.~(\ref{eq:ic}),  so to be consistent with the experimental fluid volume 
(in units of $a^3)$ by $V = 2 \pi S h_0 w$, with $S$ in 
units of $a$ and $h_0 = 1$. We note that the choice that has been made in 
selecting the parameters (in particular, having fixed film thickness and varying 
fluid volume as the opening angle $\alpha$ is modified) simplifies the connection to the 
experiments; the price to pay is the increased complexity of the results, in 
particular when discussing the trends of the results as $\alpha$ is modified,
as we will see in what follows.

Figure~\ref{fig:funnel_cv_exp} shows the results for both funnel 
simulations (marked by `F') and for the same fluid volume travelling 
down an incline plane (marked by  `P'), for the fluid volumes corresponding
to the experimental ones.  When comparing the thicknesses of the capillary
ridge between funnel and incline plane flow, the effects related to convergent
nature of the funnel flow become relevant, as discussed in Sec.~\ref{sec:numerical}.  
In particular, when considering the change of the capillary ridge thickness as the opening 
angle $\alpha$ is varied, we need to remember that in the simulations 
the volume increases as $\alpha$ decreases from $60^\circ$ 
to $45^\circ$ and $30^\circ$ (to keep approximate consistency
with the experiments), leading to thicker films and ridges. ~\footnote{Since we are interested
more in the trends than in exactly matching theory with experiments, we keep round numbers for the angles that we use, since 
the differences in the results are minor.}
The influence of the volume increase is visible in figure~\ref{fig:funnel_cv_exp}, 
where we observe non-monotonous dependence of the  capillary ridge thickness on $\alpha$.  

Next we proceed with application of the LSA to the present problem.  
To make progress, we choose an approach that allows us to reach a basic understanding of the instability 
development observed in the experiments.  The LSA, as already discussed, is based
on the incline plane problem and the constant flux setup, using the film 
thickness behind the capillary ridge, $h_0$, as the appropriate scale, see figure~\ref{fig:LSA}.  
We assume that the film is initially deposited at $r= r_{c0}$, so that the fluid volume forms 
a circle of radius $r_{c0}\cos\alpha$. As the film flows down a funnel, the radius of this circle, 
$r_c(t)$, becomes smaller, and the film itself thins (for the chosen initial condition).  
To make a comparison with experiments, we choose the characteristic thicknesses, $h_0$, 
as the thickness obtained from simulations at the time when the film front reaches $r_{c1}$, where 
the onset of fingering 
extracted from the experimental results occurs, as illustrated in figure~\ref{fig:funnel_cv_exp}.  
Table~\ref{table:prediction} lists the values of $h_0 (r_{c1})$ for a few values of $\alpha$ and for the widths $w$ of the 
initial condition that lead to the experimental fluid volumes. Additional
simulations (not shown for brevity) show that $h_0 (r_{c1})$ is essentially the same for any 
reasonable choice of the initial fluid geometry. 

\begin{figure}
\centering
\includegraphics[scale=1.0]{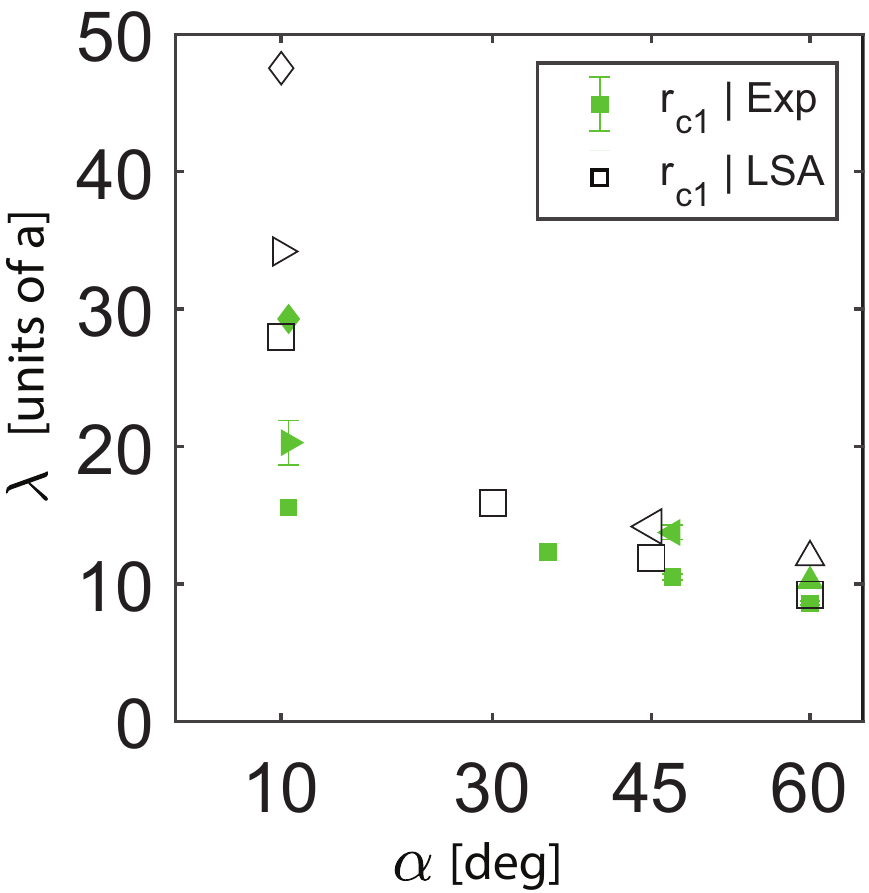}
\caption{(Color online) Funnel: average finger spacing wavelength, $\lambda$, versus opening 
angle, $\alpha$, including  experiments (Exp, filled green symbols) for which $r_{c1}$ is available (see Table~\ref{table:exp}), as well as matching linear stability analysis (LSA, open black symbols) predictions from Table~\ref{table:prediction}. 
For completeness we plot the results for  all available values of $h_i$ per angle. Symbols indicate $h_i$. $\square$: $h_i = 5$ mm for $60^\circ,~47^\circ,~35^\circ$, and $h_i = 2.5$ mm for $11^\circ$; $\bigtriangleup$: $h_i = 7$ mm for $60^\circ$; $\triangleleft$: $h_i = 7$ mm for $47^\circ$; $\triangleright$: $h_i = 3$ mm for $11^\circ$; $\diamond$: $h_i = 4$ mm for $11^\circ$}
\label{fig:experiment_data_quant}
\end{figure}

Figure~\ref{fig:experiment_data_quant} plots the obtained results for the most unstable wavelength predicted
by the LSA together with the experimentally measured one; the LSA results are also shown in Table~\ref{table:prediction}.  
The experimental wavelength is defined by
$\lambda = 2\pi\cos(\alpha)r_{c}/N_{\rm exp}$, and is given in units of the capillary length.  
We note that it is crucial to use the film thickness $h_0(r_{c1})$ when 
comparing the predictions of the LSA and experiment: using the initial 
film thickness does not lead to a meaningful agreement.  
The number of fingers predicted by the LSA, $N_{\rm LSA}$ (given also in Table~\ref{table:prediction}), 
can be compared directly with the values obtained in the experiments, $N_{\rm exp}$, see Table~\ref{table:exp}.
We find that the agreement is excellent for larger opening angles, however for small angles the 
most unstable wavelength found by the LSA is larger than the experimental one.  There is a number
of possible reasons for this difference, including an increased influence of the azimuthal curvature
that is not included in the presented methodology 
(note that $r_{c1}$ is smaller for small $\alpha$'s), or 
simply the fact that slow development of instability for small values of $\alpha$ may involve
additional effects, such as transient growth mechanism which was proposed for the
flow down an incline~\citep{BB97}.

Next, we proceed with explanation (at least in qualitative terms) of some perhaps counterintuitive trends of the 
experimental results.  Figure~\ref{fig:experiment_data_fingers}(d) shows an {\it increase} of the finger
tip speed as the opening angle is {\it decreased}.  Measuring the typical slopes in this figure, we find that 
finger tip speed increases by the factors of (approximately) $1.5$ and $2.5$ as the
opening angle decreases from $\alpha = 60^\circ$ to $47^\circ$ and $35^\circ$, respectively.   
Recalling now the expected scaling for the front speed, $U \sim h_0^2 \sin \alpha$, 
and using the values for $h_0$ at the instability onset from 
Table~\ref{table:prediction}, we find that this expression for $U$ provides a good 
approximation for the front tip speed (the corresponding ratios are approximately
1.8 and 2.2).  Therefore, an increase of the value of $h_0$ as the opening angle decreases has stronger
influence then a decrease of $\sin(\alpha)$.  
One should keep in mind of course that the above expression for $U$ 
applies to an unperturbed front while the experimental results from figure~\ref{fig:experiment_data_fingers} are obtained by measuring
the tip speed, and therefore only approximate agreement could be expected.   

To explain an increase of the typical instability wavelength as $\alpha$ is decreased, recall that based on the 
standard scaling argument, the (dimensional) most unstable wavelength scales with the film thickness 
behind the front, $h_0$, see Appendix ~\ref{app:incline}. 
Note that this scaling argument is approximate only, 
since based on the LSA for the constant flux flow down an incline, see figure~\ref{fig:LSA_D} in Appendix~\ref{app:incline}, 
the inclination angle influences the most unstable wavelength as well.  Still, the scaling $\lambda_m \propto h_0 (r_{c1})$ appears to be a good description of the experimental results, 
as it can be seen from Tables~\ref{table:exp} and~\ref{table:prediction}.

\begin{table}
\centering
\begin{tabular}{>{\centering}m{2.0cm}| c| >{\centering}m{1.5cm} | >{\centering}m{1.5cm} || >{\centering}m{1cm} | >{\centering}m{1cm} | >{\centering}m{1cm} |c}
 $\alpha$ (degrees) & $w$ & $r_{c1}$ & $h_0 (r_{c1})$ & $q_m$ & $q_c$ & $\lambda_m$ & $N_{\rm LSA}$ \\
 \hline
\multirow{2}{*}{60} & \multirow{1}{*}{3.3} 
  &  73.0 & 0.3 & 0.68 & 1.15 & 9.24 & 25 \\
 & \multirow{1}{*}{6.4} 
  & 74.7 & 0.6 & 0.52 & 0.89 & 11.97 & 20 \\
 \hline
\multirow{2}{*}{45} & \multirow{1}{*}{5.3} 
&  50.0 & 0.45 & 0.53 & 0.89 & 11.91 & 19 \\
 & \multirow{1}{*}{10.5} 
 &  48.0 & 0.69 & 0.44 & 0.75 & 14.19 & 15 \\
 \hline
\multirow{1}{*}{30} & \multirow{1}{*}{8.3} 
 & 38.5 & 0.59 & 0.40 & 0.67 & 15.90 & 13 \\
\hline
\multirow{2}{*}{10} & \multirow{1}{*}{7.9} 
 &  30.7 & 0.56  & 0.22 & 0.37 & 27.97 & 6 \\
 & \multirow{1}{*}{11.4} 
&  33.3 & 0.73  & 0.18 & 0.29 & 34.20 & 6 \\
 & \multirow{1}{*}{21.0} 
 &  28.7 & 0.99  & 0.13 & 0.21 & 47.53 & 3 \\
\hline
\end{tabular}
\caption{Results and predictions of the linear stability analysis for the experimental fluid
volumes.  The initial film thickness $h_0 (r_{c0}) = 1.0$.  The columns are as follows:
$\alpha$: the funnel opening angle, similar to the experimental values, see Table~\ref{table:exp}; 
$w$: the width of the initial condition in time-dependent simulations as used in figure~\ref{fig:funnel_cv_exp}, 
$r_{c1}$: the position at which instability is observed in the experiments;
$h_0$ at $r_{c1}$, as obtained in the simulations for funnel flow, see the text for details; 
$q_m$: the most unstable wavenumber obtained by the LSA;
$q_c$: critical wavenumber obtained by the LSA; 
$\lambda_m = {2 \pi/q_m}$; 
$N_{\rm LSA}$: prediction for the  number of fingers based on $\lambda_m$ and $r_{c1}$.
The  values used for  $r_{c0}$ (initial front position) and coming from the 
experiments and are listed in Table~\ref{table:exp};  note that the listed values
of $w$ combined with the specified values of $r_{c0}$ and $h_0 (r_{c0})$ lead to the same fluid volume as in the experiments.
}
\label{table:prediction}
\end{table}

\subsubsection{Discussion}

Reasonable agreement between theory and experiments, in particular for larger opening
angles, shows that our approach combining the information from 
experiments, simulations, and LSA describes well the main features of instability development. 
Before closing this section, we list few additional comments and observations.

\begin{itemize}
    \item 
    The reader may wonder whether simply using $h_i$ from the experiments (see Table~\ref{table:exp})
    could be used to describe the instability development and emerging lengthscales.  This 
    approach however does not lead to a reasonable estimate, since there is a considerable change in 
    the film thickness between $r_{c0}$ and $r_{c1}$.   One may also wonder whether the results of LSA 
    may depend on the initial condition in simulations (which is chosen ad hoc): the  answer is again 
    no, since the film does not develop instability immediately; by the time $r_{c1}$ is reached, 
    the memory of the initial condition is lost. 
    \item
There are two main differences between the flow down an incline and in a funnel: one of them is 
film thickening due to convergent nature of the flow, as discussed in Sec.~\ref{sec:self-similar} 
and~\ref{sec:numerical}; the other one is the presence of azimuthal curvature for the funnel flow,
that we have not discussed in much detail.   The curvature in the azimuthal direction scales as 
$1/r$, see Section~\ref{sec:model}, so it is a small quantity as long as only large values of  $r$ 
are considered.  This value should be compared with the typical curvature (in the radial direction) 
of the film itself, which is an O(1) quantity close to the film front. Smallness of the azimuthal 
curvature justifies ignoring it in the present work, since it is expected to become important only 
very close to the funnel centre.  Therefore, as long as the fluid front is far away from the centre, 
the flow in a funnel is similar to the flow down an incline plane, as long as the fact that the film thickens due 
to volume conservation is taken into account.
\item 
Figure~\ref{fig:funnel_cv_exp} specifies the times at which instability starts to develop 
(when the fluid front reaches $r_{c1}$).  We note that these times are shorter for the funnel compared to flow down an incline,
in particular for smaller $\alpha$'s; this is due to the film
thickening for the flow in a funnel.  Thicker films flow faster and also become unstable sooner, 
compared to the flow down an incline.
\item 
In light of the discussion in this section, the experimental fact that the observed number of fingers 
does not depend on fluid viscosity (see \textit{Supplementary Table 1}) may not be obvious.  While viscosity only changes the time 
scale of the flow, for the present problem the time scale may be important since the film thickness 
changes with time.  However, the location at which film becomes unstable, $r_{c1}$, and the film 
thickness behind the capillary ridge, $h_0$, turn out not to depend on the fluid viscosity, supporting 
the presented approach for carrying out the LSA and interpretation of the results.
\item
The LSA predicts that instability will develop if the circumference $2\pi r_{c}\cos\alpha$ is 
larger than $\lambda_c = {2 \pi/q_c}$.  
Consistently, the maximum number of unstable modes (leading to fingers in experiments) that can be 
supported is $\floor{2\pi r_{c}\cos\alpha/\lambda_c}$, where $\floor{\cdot}$ is the floor function.  
One untested consequence of this result is that if the film is released close to the centre of the funnel, 
it may not become unstable since the circumference of the circle formed by the initial fluid front 
may not be long enough to support instability development.  
\end{itemize}

\section{Conclusions}\label{sec:conclusions}

The presented results show that a reasonably complete understanding of instability development 
for a film flowing in a funnel 
can be reached by combining the insight from experiments and asymptotic analysis 
that allows for significant simplification of the governing equations.  Furthermore, it turns
out that despite the complexity of the problem, a useful insight can be also reached by considering
a self-similar approach similar to the one used for the flow down an incline plane.  Such insight from 
self-similar methods combined with linear stability analysis originated from the flow down an incline provides an important guidance in carrying
out numerical simulations that help to develop better understanding of the instability development. 
While we have focused on a particular geometry of flow in a funnel, we note that a similar approach could be 
applied to a number of other unstable flows, such as the flows on a sphere, outside surface of a 
funnel, or even in more complicated geometries.

To conclude, we note that instabilities of the systems whose base state evolves in time are difficult to analyze in 
a tractable manner.  For the present problem, we have shown that a reasonably good insight
can be reached by simplifying the problem first, and then using some input regarding 
instability development from the experiments.  One would of course like to be able to 
understand the general features of instability development, including the factors that 
govern instability onset itself.   Reaching this goal will require further development 
of stability analysis and is left as an open problem for future work.

{\it Acknowledgements} We thank Vittorio Saggiomo for help with 3D printing of the funnels, to 
the undergraduate students Elliot Figueroa, Jody Parchment and Yimei Xu (supervised by Ryan Allaire) for 
help with the experiments and data analysis as a part of their Capstone class at NJIT. 
TSL acknowledges support by Ministry of Science and Technology, Taiwan, under research grant MOST-109-2115-M-009-006-MY2, 
and LK by the NSF grants CBET-1604351 and DMS-1815613.  

\appendix

\section{Stability of the constant flux flow down an incline plane \label{app:incline}}

\begin{figure}
\centering
\includegraphics[scale=0.4]{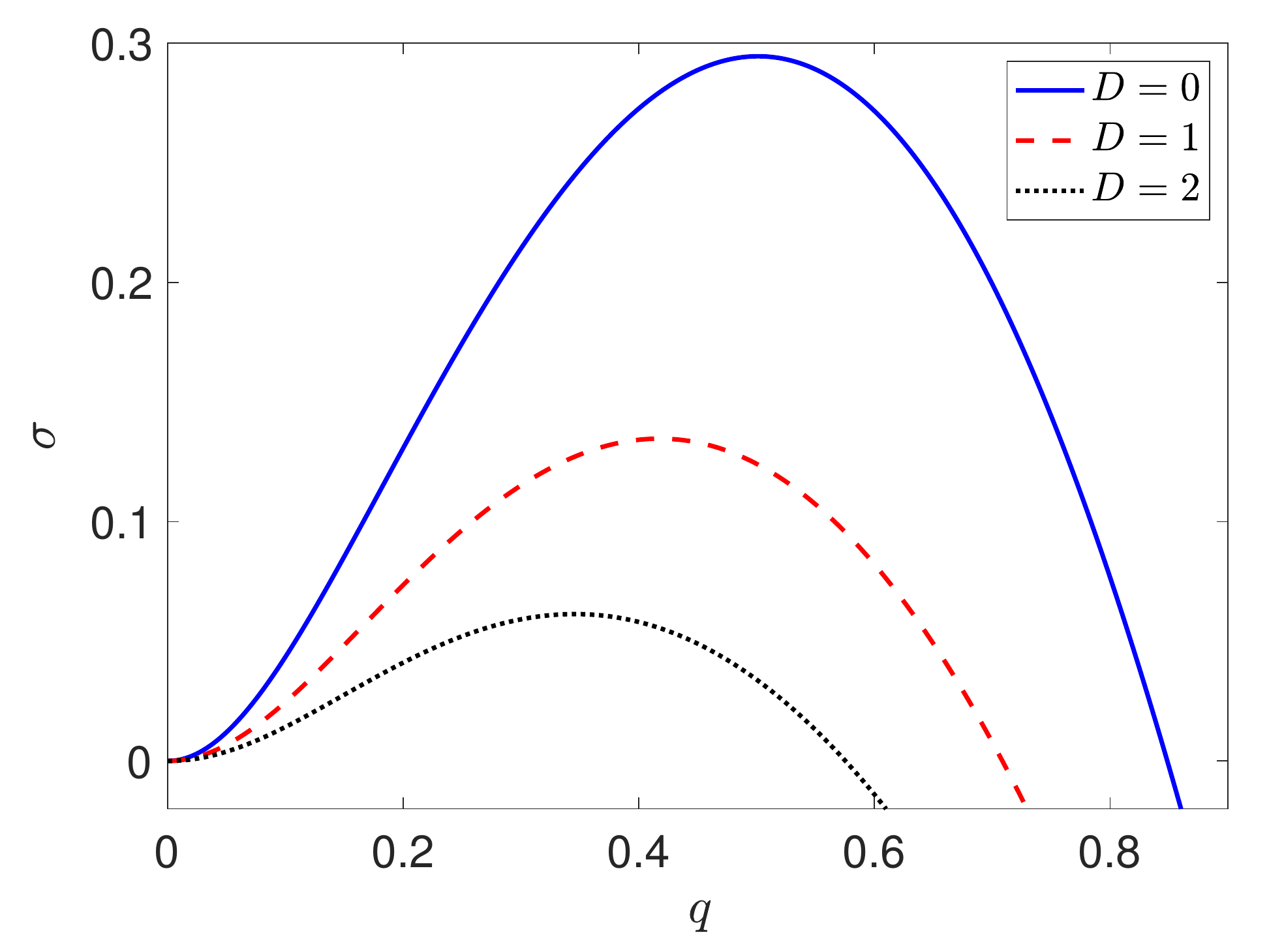}
\caption{(Color online) 
Results of LSA for liquid film on an incline plane with different $D$. The precursor film thickness is $b_0=0.01$. 
}
\label{fig:LSA_D}
\end{figure}

Consider a completely wetting fluid flowing down a planar surface enclosing an angle $\alpha$ with the horizontal. With the 
same scales as used in the main body of the paper,
the evolution equation of the film thickness can be written as (see, e.g., ~\citep{BB97, Kondic2003})
\begin{equation}
\frac{\partial}{\partial t} h = -\nabla\cdot\left[h^3\,\nabla^3 h - \cos\alpha h^3\nabla h +\sin\alpha h^3 \vect{e}_x\right],
\label{eq_x}
\end{equation}
where $\vect{e}_x$ is the unit vector pointing in the positive $x$-direction. 
Re-normalizing the variables as
$
h = h_0 \bar{h},~ t = \bar{t}/{ (h_0^{5} \sin^{4}\alpha)^{1/3}},~ x = ({h_0/\sin\alpha})^{1/3} \bar{x},
$
one obtains the well-known model of thin liquid film (after dropping the bars)
\begin{equation}
\frac{\partial}{\partial t} h = -\nabla\cdot\left[h^3\,\nabla^3 h - D h^3\nabla h +h^3 \vect{e}_x\right],
\end{equation}
where $D= (h_0/\sin\alpha)^{2/3}\cos\alpha$. This equation admits a one-dimensional travelling wave solution satisfying
\begin{equation}
-Uh + h^3\,h_{xxx} - D h^3\,h_x + h^3=c,
\label{eq_TW}
\end{equation}
where boundary conditions $h(x=\infty)=b_0$, $h(x=-\infty)=1$ are imposed to have 
$U = 1+b_0+b_0^2$, $c = -(b_0+b_0^2).$ 
(Note that the (unscaled) speed of the front scales with the film thickness squared, as it can be seen from 
the ratio of the scaling factors for $x$ and $t$.) 
Linear stability of the film with respect to perturbations in the transverse, $y$, direction is conveniently carried out in  
a moving coordinate frame, $s = x - Ut$,  where we assume the solution of the form
\begin{equation}
h(s, y, t) = H(s) + \epsilon g(s)e^{\sigma t} e^{i q y},
\label{eq_ansatz_tw}
\end{equation}
where $H(s)$ satisfies (\ref{eq_TW}).  
At $O(\epsilon)$ we obtain a linear eigenvalue problem for $g(s; q)$ with eigenvalue $\sigma$ that represents 
the growth rate of temporal evolution of the perturbation at each wave number. 
Figure~\ref{fig:LSA_D} shows the results of this analysis; for more details see e.g.~\citet{Kondic2003}.

\bibliographystyle{jfm}
\bibliography{films.bib}

\begin{thebibliography}{40}
\expandafter\ifx\csname natexlab\endcsname\relax\def\natexlab#1{#1}\fi

\bibitem[Backholm {\em et~al.\/}(2014)Backholm, Benzaquen, Salez, Raphaël \&
  Dalnoki-Veress]{2014_SM_capillary_collapse}
{\sc Backholm, M., Benzaquen, M., Salez, T., Raphaël, E. \& Dalnoki-Veress,
  K.} 2014 Capillary levelling of a cylindrical hole in a viscous film. {\em
  Soft Matter\/} {\bf 10}, 2550--2558.

\bibitem[Balestra {\em et~al.\/}(2019)Balestra, Badaoui, Ducimetière \&
  Gallaire]{balestra_jfm_2019}
{\sc Balestra, G., Badaoui, M., Ducimetière, Y.-M. \& Gallaire, F.} 2019
  Fingering instability on curved substrates: optimal initial film and
  substrate perturbations. {\em J. Fluid Mech.\/} {\bf 868}, 726.

\bibitem[Bertozzi \& Brenner(1997)]{BB97}
{\sc Bertozzi, A.~L. \& Brenner, M.~P.} 1997 {Linear stability and transient
  growth in driven contact lines}. {\em Phys. Fluids\/} {\bf 9}, 530.

\bibitem[Bonn {\em et~al.\/}(2009)Bonn, Eggers, Indekeu, Meunier \&
  Rolley]{bonn_rmp2009}
{\sc Bonn, D., Eggers, J., Indekeu, J., Meunier, J. \& Rolley, E.} 2009
  {Wetting and spreading}. {\em Rev. Mod. Phys.\/} {\bf 81}, 739.

\bibitem[Bostwick {\em et~al.\/}(2017)Bostwick, Dijksman \&
  Shearer]{bostwick_prf_2017}
{\sc Bostwick, J.~B., Dijksman, J.~A. \& Shearer, M.} 2017 Wetting dynamics of
  a collapsing fluid hole. {\em Phys. Rev. Fluids\/} {\bf 2}, 014006.

\bibitem[Brand\~ao {\em et~al.\/}(2014)Brand\~ao, Fontana \&
  Miranda]{miranda_pre_2014}
{\sc Brand\~ao, R., Fontana, J.~V. \& Miranda, J.~A.} 2014 Suppression of
  viscous fingering in nonflat hele-shaw cells. {\em Phys. Rev. E\/} {\bf 90},
  053003.

\bibitem[Craster \& Matar(2009)]{cm_rmp09}
{\sc Craster, R.V. \& Matar, O.K.} 2009 Dynamics and stability of thin liquid
  films. {\em Rev. Mod. Phys.\/} {\bf 81}, 1131.

\bibitem[Davis(1980)]{davis_jfm80}
{\sc Davis, S.~H.} 1980 {Moving contact lines and rivulet instabilities. {P}art
  I: The static rivulet}. {\em J. Fluid Mech.\/} {\bf 98}, 225.

\bibitem[De~Gennes {\em et~al.\/}(2004)De~Gennes, Brochard-Wyart \&
  Qu{\'e}r{\'e}]{degennes2004capillarity}
{\sc De~Gennes, Pierre-Gilles, Brochard-Wyart, Fran{\c{c}}oise \&
  Qu{\'e}r{\'e}, David} 2004 {\em Capillarity and wetting phenomena: drops,
  bubbles, pearls, waves\/}. Springer Science \& Business Media.

\bibitem[Diez {\em et~al.\/}(2009)Diez, Gonz{\'a}lez \& Kondic]{dgk_pof09}
{\sc Diez, J.~A., Gonz{\'a}lez, A. \& Kondic, L.} 2009 {On the breakup of fluid
  rivulets}. {\em Phys. Fluids\/} {\bf 21}, 082105.

\bibitem[Diez {\em et~al.\/}(1992)Diez, Gratton \& Gratton]{DGG}
{\sc Diez, J.~A., Gratton, J. \& Gratton, J.} 1992 {Self-similar solutions of
  the second kind for a convergent viscous gravity current}. {\em Phys. Fluids
  A\/} {\bf 4}, 1148.

\bibitem[Diez {\em et~al.\/}(2001)Diez, Kondic \& Bertozzi]{DKB01}
{\sc Diez, J.~A., Kondic, L. \& Bertozzi, A.~L.} 2001 {Global models for moving
  contact lines}. {\em Phys. Rev. E\/} {\bf 63}, 011208.

\bibitem[Dijksman {\em et~al.\/}(2019)Dijksman, Mukhopadhyay, Behringer \&
  Witelski]{dijksman2019}
{\sc Dijksman, J.~A., Mukhopadhyay, S., Behringer, R.~P. \& Witelski, T.~P.}
  2019 Thermal {M}arangoni-driven dynamics of spinning liquid films. {\em Phys.
  Rev. Fluids\/} {\bf 4}, 084103.

\bibitem[Dukler {\em et~al.\/}(2020)Dukler, Ji, Falcon \& Bertozzi]{DJFB20}
{\sc Dukler, Y., Ji, H., Falcon, C. \& Bertozzi, A.~L.} 2020 {Theory for
  undercompressive shocks in tears of wine}. {\em Phys. Rev. Fluids\/} {\bf 5},
  034002.

\bibitem[Fraysse \& Homsy(1994)]{FH94}
{\sc Fraysse, N. \& Homsy, G.~M.} 1994 {An experimental study of rivulet
  instabilities in centrifugal spin coating of viscous \mbox{Newtonian} and
  \mbox{non-Newtonian} fluids}. {\em Phys. Fluids\/} {\bf 6}, 1491.

\bibitem[Gomba {\em et~al.\/}(2007)Gomba, Diez, Gratton, Gonzalez \&
  Kondic]{Gomba2007}
{\sc Gomba, J.~M., Diez, J., Gratton, R., Gonzalez, A.~G. \& Kondic, L.} 2007
  {Stability study of a constant-volume thin film flow}. {\em Phys. Rev. E\/}
  {\bf 76}, 046308.

\bibitem[Gonzalez {\em et~al.\/}(2013)Gonzalez, Diez \& Kondic]{gdk_jfm13}
{\sc Gonzalez, A.~G., Diez, J.~D. \& Kondic, L.} 2013 Stability of a liquid
  ring on a substrate. {\em J. Fluid Mech.\/} {\bf 718}, 213.

\bibitem[Goodwin \& Homsy(1991)]{hom91}
{\sc Goodwin, R. \& Homsy, G.~M.} 1991 {Viscous Flow down a slope in the
  vicinity of a contact line}. {\em Phys. Fluids A.\/} {\bf 3}, 515.

\bibitem[Ho \& Khew(2000)]{2000_Ho_latex}
{\sc Ho, C.~C. \& Khew, M.~C.} 2000 Surface free energy analysis of natural and
  modified natural rubber latex films by contact angle method. {\em Langmuir\/}
  {\bf 16}, 1407--1414.

\bibitem[Howell(2003)]{Howell2003}
{\sc Howell, P.~D.} 2003 {Surface tension driven flow on a moving curved
  surface}. {\em J. Eng. Math.\/} {\bf 45}, 283--308.

\bibitem[Huppert(1982)]{Hupp82}
{\sc Huppert, H.} 1982 {Flow and Instability of a viscous current down a
  slope}. {\em Nature\/} {\bf 300}, 427.

\bibitem[Kondic(2003)]{Kondic2003}
{\sc Kondic, L.} 2003 Instabilities in gravity driven flow of thin fluid films.
  {\em SIAM Review\/} {\bf 45}, 95.

\bibitem[Kondic(2019)]{web_capstone}
{\sc Kondic, L.} 2019 {Capstone Laboratory}.
  {http://cfsm.njit.edu/capstone/projects/2019/main.php}.

\bibitem[Lin {\em et~al.\/}(2020)Lin, Dijksman \& Kondic]{SM}
{\sc Lin, T.-S., Dijksman, J. \& Kondic, L.} 2020 See supplementary material.

\bibitem[Lin \& Kondic(2010)]{Lin2010}
{\sc Lin, T.-S. \& Kondic, L.} 2010 {Thin films flowing down inverted
  substrates: Two dimensional flow}. {\em Phys. Fluids\/} {\bf 22}, 052105.

\bibitem[Lv {\em et~al.\/}(2018)Lv, Eigenbrod \& Hardt]{hardt_jfm_2018}
{\sc Lv, C., Eigenbrod, M. \& Hardt, S.} 2018 Stability and collapse of holes
  in liquid layers. {\em J. Fluid Mech.\/} {\bf 855}, 1130–1155.

\bibitem[Marchand {\em et~al.\/}(2012)Marchand, Das, Snoeijer \&
  Andreotti]{2012_prl_elastocap}
{\sc Marchand, A., Das, S., Snoeijer, J.~H. \& Andreotti, B.} 2012 Contact
  angles on a soft solid: {F}rom {Y}oung's {L}aw to {N}eumann's {L}aw. {\em
  Phys. Rev. Lett.\/} {\bf 109}, 236101.

\bibitem[Mayo {\em et~al.\/}(2013)Mayo, McCue \& Moroney]{Mayo2013}
{\sc Mayo, L.~C., McCue, S.~W. \& Moroney, T.~J.} 2013 Gravity-driven fingering
  simulations for a thin liquid film flowing down the outside of a vertical
  cylinder. {\em Phys. Rev. E\/} {\bf 87}, 053018.

\bibitem[Melo {\em et~al.\/}(1989)Melo, Joanny \& Fauve]{Melo}
{\sc Melo, F., Joanny, J.~F. \& Fauve, S.} 1989 {Fingering Instability of
  Spinning Drops}. {\em Phys. Rev. Lett.\/} {\bf 63}, 1958.

\bibitem[Miranda {\em et~al.\/}(2000)Miranda, Parisio, Moraes \&
  Widom]{miranda_pre_2000}
{\sc Miranda, J.~A., Parisio, F., Moraes, F. \& Widom, M.} 2000 Gravity-driven
  instability in a spherical hele-shaw cell. {\em Phys. Rev. E\/} {\bf 63},
  016311.

\bibitem[Oron {\em et~al.\/}(1997)Oron, Davis \& Bankoff]{oron_rmp97}
{\sc Oron, A., Davis, S.~H. \& Bankoff, S.~G.} 1997 {Long-scale evolution of
  thin liquid films}. {\em Rev. Mod. Phys.\/} {\bf 69}, 931.

\bibitem[Pismen \& Eggers(2008)]{Pismen2008}
{\sc Pismen, L.~M. \& Eggers, J.} 2008 Solvability condition for the moving
  contact line. {\em Phys. Rev. E\/} {\bf 78}, 056304.

\bibitem[Roy {\em et~al.\/}(1997)Roy, Roberts \& Simpson]{Roy1997}
{\sc Roy, R.~V., Roberts, A.~J. \& Simpson, M.~E.} 1997 {A lubrication model of
  coating flows over a curved substrate in space}. {\em J. Fluid Mech.\/} {\bf
  454}, 235--261.

\bibitem[Sibley {\em et~al.\/}(2015)Sibley, Nold \& Kalliadasis]{Sibley2015}
{\sc Sibley, D., Nold, A. \& Kalliadasis, S.} 2015 The asymptotics of the
  moving contact line: cracking an old nut. {\em J. Fluid Mech.\/} {\bf 764},
  445--462.

\bibitem[Smolka \& SeGall(2011)]{Smolka2011}
{\sc Smolka, L. \& SeGall, M.} 2011 {Fingering instability down the outside of
  a vertical cylinder}. {\em Phys. Fluids\/} {\bf 23}, 092103.

\bibitem[Snoeijer \& Andreotti(2013)]{Snoeijer2013}
{\sc Snoeijer, J.~H. \& Andreotti, B.} 2013 Moving contact lines: Scales,
  regimes, and dynamical transitions. {\em Annu. Rev. Fluid Mech.\/} {\bf 45},
  269--292.

\bibitem[Takagi \& Huppert(2010)]{huppert_jfm_2010}
{\sc Takagi, D. \& Huppert, H.~E.} 2010 Flow and instability of thin films on a
  cylinder and sphere. {\em J. Fluid. Mech.\/} {\bf 647}, 221--238.

\bibitem[Troian {\em et~al.\/}(1989)Troian, Herbolzheimer, Safran \&
  Joanny]{Troian}
{\sc Troian, S.~M., Herbolzheimer, E., Safran, S.~A. \& Joanny, J.~F.} 1989
  {Fingering Instabilities of Driven Spreading Films}. {\em Europhys. Lett.\/}
  {\bf 10}, 25.

\bibitem[Zhang {\em et~al.\/}(2018)Zhang, Vandaele, Seveno \&
  Coninck]{Zhang_2018}
{\sc Zhang, Y., Vandaele, A., Seveno, D. \& Coninck, J.~De} 2018 Wetting
  dynamics of polydimethylsiloxane mixtures on a poly(ethylene terephthalate)
  fiber. {\em J. Coll. Interface Sci.\/} {\bf 525}, 243--250.

\bibitem[Zheng {\em et~al.\/}(2018)Zheng, Fontelos, Shin, Dallaston, Tseluiko,
  Kalliadasis \& Stone]{zheng_jfm_2018}
{\sc Zheng, Z., Fontelos, M.~A., Shin, S., Dallaston, M.~C., Tseluiko, D.,
  Kalliadasis, S. \& Stone, H.~A.} 2018 Healing capillary films. {\em J. Fluid
  Mech.\/} {\bf 838}, 404--434.

\end{thebibliography}

\end{document}